\documentclass[journal]{IEEEtran}
\usepackage{cite}
\usepackage{svg}
\usepackage{amsmath} 
\usepackage{float}
\usepackage{subfigure}
\usepackage[]{graphicx}
\usepackage{amsmath}
\usepackage{cases}
\usepackage{stfloats}
\usepackage{amsfonts}
\usepackage{subeqnarray}
\usepackage{longtable}
\usepackage{supertabular}
\usepackage{setspace}
\usepackage{multirow}
\usepackage{booktabs}
\usepackage[ruled,linesnumbered]{algorithm2e}
\makeatletter
\newcommand{\nosemic}{\renewcommand{\@endalgocfline}{\relax}}
\newcommand{\dosemic}{\renewcommand{\@endalgocfline}{\algocf@endline}}
\let\oldnl\nl
\newcommand{\nonl}{\renewcommand{\nl}{\let\nl\oldnl}}
\makeatother
\usepackage[export]{adjustbox} 
\usepackage{booktabs}
\usepackage{setspace}
\usepackage{xcolor}
\usepackage{mathrsfs}
\usepackage{amsmath}
\usepackage{array}
\usepackage{amssymb}
\usepackage{amsthm}
\usepackage{microtype}
\usepackage{url}
\usepackage{amsfonts,amssymb}
\usepackage{dsfont}
\usepackage{mathtools}
\usepackage{xcolor,colortbl}
\usepackage{colortbl}
\usepackage{graphicx}

\definecolor{Gray}{gray}{0.85}
\definecolor{Whitecolor}{rgb}{1,1,1}

\newcolumntype{a}{>{\columncolor{Gray}}c}
\newcolumntype{b}{>{\columncolor{white}}c}

\allowdisplaybreaks
\begin{document}
\setstretch{1}
\title{\textls[-25]{Data-driven Emergency Frequency Control for Multi-Infeed Hybrid AC-DC System}}
\author{Qianni~Cao,
Chen~Shen,~\IEEEmembership{Senior~Member,~IEEE},
Ye~Liu,~\IEEEmembership{Student~Member,~IEEE}

\thanks{This work is supported by The National Key R\&D Program of China (2021YFB2400800). (Corresponding author: Chen Shen.)}     

\thanks{Qianni Cao, Chen Shen, and Ye Liu are with the State Key Laboratory of Power Systems, Department of Electrical Engineering, Tsinghua University, Beijing 100084, China (e-mail: cqn20@mails.tsinghua.edu.cn; shenchen@mail.tsinghua.edu.cn; liuye18@mails.tsinghua.edu.cn).}
}
        


\maketitle

\begin{abstract}
With the continuous development of large-scale complex hybrid AC-DC grids, the fast adjustability of HVDC systems is required by the grid to provide frequency regulation services. This paper develops a fully data-driven linear quadratic regulator (LQR) for the HVDC to provide temporal frequency support. The main technical challenge is the complexity and the nonlinearity of multi-infeed hybrid AC-DC (MIDC) systems dynamics that make the LQR intractable. Based on Koopman operator (KO) theory, a Koopman eigenpairs construction method is developed to fit a global linear dynamic model of MIDC systems. Once globally linear representation of uncontrolled system dynamics is obtained offline, the control term is constituted by the gradient of the identified eigenfunctions and the control matrix $\boldsymbol{B}$. In case that $\boldsymbol{B}$ is unknown, we propose a method to identify it based on the verified Koopman eigenfunctions. The active power reference is optimized online for LCC-HVDC in a moving horizon fashion to provide frequency support, with only locally measurable frequency and transmission power. The robustness of the proposed control method against approximation errors of the linear representation in eigenfunction coordinates is analyzed. Simulation results show the effectiveness, robustness and adaptability of the proposed emergency control strategy.

\end{abstract}
\IEEEpeerreviewmaketitle



\section{Introduction}
\subsection{Motivation}
\IEEEPARstart{T}{he} technical and economic advantages of HVDC transmission technologies have promoted the development of multi-infeed hybrid AC-DC (MIDC) systems \cite{2019Assessing}, in which multiple line-commutated-converter-based HVDC (LCC-HVDC) systems are connected to one AC system. In recent years, the consequent asynchronous interconnected regional power grids, complicated system dynamics and possible emergency faults of MIDC systems pose serious threats to the frequency stability of the system \cite{2021Optimal}. Frequency stability issues are caused by power imbalance. To deal with the considerable power imbalance in MIDC systems, an emergency frequency control (EFC) strategy is indispensable. Apart from generator tripping or load shedding operations, effective EFC strategies could be designed by utilizing the fast adjustability of HVDC systems, which have the potential to improve the system frequency stability \cite{2012Integrated,9215149}. 

Considering the EFC of hybrid AC-DC systems, emergency DC power support (EDCPS) is an effective approach. In this paper, to design a decentralized approach for EFC with LCC-HVDC systems participating, a fully data-driven decentralized EFC strategy is proposed to regulate DC power reference. The implementation of this approach relies only on measurements, making it suitable for complicated MIDC systems.
\vspace{-0.2cm}
\subsection{Literature Review}
Numerous model-based EFC strategies have been developed for hybrid AC-DC systems. Ref.\cite{2021Optimal} proposed a decentralized EFC strategy based on coordinated droop for MIDC systems. It designed the optimal droop for power allocation based on state model of the system. By studying the overload capacity based on transient IGBT thermal models, Ref.\cite{2017Dynamic} investigated a frequency-power droop controller and a maximum power release controller in modular multilevel converter (MMC)-based VSC HVDC system. Ref. \cite{2019Continuous} developed a continuous under-frequency load shedding scheme and improved the scheme by analyzing an extended system frequency response (SFR) model including frequency threshold and time delay. A centralized response-based AC-DC coordinated control strategy was proposed in Ref.\cite{2015Coordinated} that combined the EDCPS strategy and load shedding operations. However, there are two main barriers hindering the application of these model-based methods. First, it is difficult to maintain the accurate models of MIDC systems. Second, the dynamic procedure of emergency faults of MIDC systems features strong nonlinearity, and the solution of a nonlinear optimization problem is not off-the-shelf.

Consequently, data-driven EFC based on measurements is showing great potential. Among the promising advances in theory and numerical approximation in data-driven control, Koopman spectral theory \cite{koopman1931hamiltonian,koopman1932dynamical} has emerged as a dominant perspective over the past decade. In Koopman spectral theory, nonlinear dynamics are represented in terms of a Koopman operator, which is an infinite dimensional linear operator acting on the space of all possible measurement functions of the system. Finding a coordinate system as finite-dimensional approximations of the Koopman operator is a one-time upfront cost for the use of highly efficient linear optimal control tools\cite{brunton2021modern}.
 
Dynamic mode decomposition (DMD) and its variants are one of the workhorse algorithms\cite{mezic2005spectral,rowley2009spectral,schmid2009dynamic} to approximate the Koopman operator. By using the dynamic mode decomposition with control (DMDc) method, Ref. \cite{2021Online} designed a wide area damping controller using discrete linear quadratic regulator (DLQR) to enhance the overall damping of low-frequency power system oscillations. However, DMDc estimate system dynamics with linear observables, which fail to capture the nonlinear transients of the system \cite{2017Data}. Extended DMD (EDMD) \cite{C2017Towards}, augmented with nonlinear functions of the measurements was recently used for model predictive control with promising results. Based on Koopman model predictive control (KMPC) in Ref. \cite{2016Linear}, a EDMD based stabilization controller was proposed in Ref.\cite{2018Power} for power grid transient stability. For other variants of DMD, Ref. \cite{2020Data} constructed a frequency predictor for the wind farm by a specialized DMD methods with specially designed Koopman observables. Powered by representation capabilities of the neural network, Ref. \cite{2021Deep} approximated the Koopman operator with the deep neural network and designed a energy storage unit controller to enhance transient stability. To realize distributed control with partial measurements, Ref. \cite{2019Frequency} designed Koopman observables in the form of time-delayed embeddings to damp frequency oscillations.

Despite impressive practical success of DMD and its variants, there is no guarantee that nonlinear functions of the measurements found will form a closed subspace under the Koopman operator\cite{2017Data}. Methods to directly identify eigenpairs under Koopman reduced-order nonlinear identification and control (KRONIC)\cite{2017Data}, \cite{Kaiser_2021} and Koopman canonical transform (KCT)\cite{2016Linear2} were introduced to address these issues. KRONIC and KCT promised to construct low-dimensional and closed models by model validation. Compared with DMD and its variants, the methods to directly identify eigenpairs are potential to achieve better modelling accuracy of system dynamics. In the light of the above analysis, this paper is based on verified eigenfunctions for MIDC systems to design data-driven EFC. Especially, the impacts of inevitable representation errors of eigenpairs on the controlled system dynamics are studied.

\vspace{-0.3cm}
\subsection{Contribution}
In this paper, we develop an Ensembled-Koopman-Emergency-Frequency-Control (EKEFC) strategy, which optimizes the active power reference for each LCC-HVDC in a moving horizon fashion to provide emergency frequency support. To deal with strong nonlinear system dynamics, the proposed strategy finds linear embeddings of nonlinear MIDC system dynamics based on the Koopman theory to facilitate the use of the mature optimizer LQR. EKEFC is purely data-driven because the globally linear representation of system dynamics are generated directly from history data. Moreover, it requires only the local frequency and the DC transmission power measurements as inputs, making it possible for distributed implementation.

In summary, the contributions of this paper are as follows:
\begin{enumerate}
    \item To approximate Koopman operator for frequency dynamics in MIDC systems, a Koopman eigenpairs construction method is developed. Physical knowledge of MIDC systems and a library bagging technique are introduced to power the construction method. Thus, the nonlinear MIDC system dynamics are reformulated with global accuracy in Koopman eigenfunction coordinates.
    \item A fully data-driven dynamic optimal control method, named as EKEFC, for multi-infeed hybrid AC-DC system frequency support is proposed. By combining the global linear dynamic model of frequency dynamics in MIDC systems, a fully data-driven LQR is designed.
    \item The robustness of EKEFC against approximation errors of the linear representation in eigenfunction coordinates is analyzed. Specifically, we provide a sufficient condition that guarantees the stability of MIDC systems with EKEFC when there are Koopman eigenpairs approximation errors. Furthermore, the error bound of the closed-loop dynamics with consideration of the approximation errors is estimated.
\end{enumerate}

The rest of this paper is organized as follows. Section II proposes the EKEFC strategy for MIDC systems. Section III examines the effect of an error in the representation of Koopman eigenpairs and provides a sufficient condition for the stability of MIDC systems with EKEFC. In Section IV, an MIDC system case is presented and the effectiveness of the proposed control strategy is verified. Section V provides the conclusion.

\section{Emergency Frequency Controller Design}
\label{Emergency_Frequency_Controller_Design}
In this section, the EKEFC strategy for MIDC systems is proposed. Firstly, we briefly overview Koopman operator theoretic concepts. Secondly, by introducing a library bagging technique, a data-driven modelling method is developed to reformulate nonlinear frequency dynamics in Koopman eigenfunction coordinates. Thirdly, control strategies are formulated directly in the eigenfunction coordinates. In case that the control matrix is unknown, we propose to identify it from data by combing the identified Koopman eigenpairs. To realize distributed control, we further select a special set of eigenpairs and revise them to adapt to partial measurements.
\vspace{-0.3cm}
\subsection{Preliminaries for Identifying Koopman Eigenfunctions} \label{Preliminaries_for_Identifying_Koopman_Eigenfunctions}
Koopman operator is a linear but an infinite-dimensional operator that governs the time evolution of observables or outputs defined on the state space of a dynamical system\cite{koopman1931hamiltonian}. In particular, we consider unactuated, autonomous dynamic systems of the form
\begin{align}
    \dot{\textbf{\textit{x}}}=\textbf{\textit{f}}(\textbf{\textit{x}}), \label{system_autonomous_dynamics}
\end{align}
with the state vector $\textbf{\textit{x}}\in\mathbb{R}^{n}$.

The Koopman operator is a linear operator $\mathcal{K}$ which advances a measurement function $\boldsymbol{g}(\boldsymbol{x})$ of the state forward in time through the dynamics
\begin{equation}
    (\mathcal{K}\boldsymbol{g})(\boldsymbol{x})=\boldsymbol{g}(\boldsymbol{f}(\boldsymbol{x}))
\label{eq_Koopman_operator}
\end{equation}

For an eigenfunction $\varphi$ of $\mathcal{K}$, corresponding to an eigenvalue $\lambda$, this becomes 
\begin{equation}
    (\mathcal{K}\varphi)(\boldsymbol{x})=\lambda\varphi(\boldsymbol{x})=\varphi(\boldsymbol{f}(\boldsymbol{x})),
\label{eq_Koopman_operator}
\end{equation}
where $(\lambda,\varphi)$ forms an eigenpair.

The observable vector in intrinsic Koopman eigenfunction coordinates is defined as
\begin{equation}
  \boldsymbol{\varphi}(\textbf{\textit{x}})=\left[\varphi_1(\textbf{\textit{x}}),\varphi_2(\textbf{\textit{x}}),\dots,\varphi_S(\textbf{\textit{x}})\right]^\text{T}, \label{observable_vector_in_intrinsic_Koopman_eigenfunction_coordinates}
\end{equation}
where $\boldsymbol{\varphi}=\{{\varphi }_{s}:{{\mathbb{R}}^{n}}\to \mathbb{C},s=1,2,...,S\}$ represents a nonlinear transformation of the state $\boldsymbol{x}$ into eigenfunction coordinates. If $\boldsymbol{\varphi}$ is differentiable at $\boldsymbol{x}$, by applying the chain rule its evolution equation can be written as
\begin{align}
    \dot{\boldsymbol{\varphi}}=\nabla{\boldsymbol{\varphi}}(\textbf{\textit{x}})\cdot\textbf{\textit{f}}(\textbf{\textit{x}}). \label{chain_rule}
\end{align}
According to the definition of Koopman eigenfunctions, we obtain
\begin{gather}
    \dot{\boldsymbol{\varphi}}=\boldsymbol{\Lambda}\boldsymbol{\varphi}(\textbf{\textit{x}}),   \label{eq_autonomous_dynamics}
\end{gather}
where $\boldsymbol{\Lambda}=diag(\lambda_1,...,\lambda_S)$ is a matrix, with diagonal elements consisting of the eigenvalue $\lambda_s(s=1,2,...,S)$ associated with the eigenfunction $\varphi_{s}$.

Combining Eq.\eqref{chain_rule} and Eq.\eqref{eq_autonomous_dynamics}, the following Koopman partial differential equation (PDE) should be satisfied by regression:
\begin{align}
    \nabla \boldsymbol{\varphi}(\textbf{\textit{x}}) \textbf{\textit{f}}(\textbf{\textit{x}})\text{=}\boldsymbol{\Lambda }\boldsymbol{\varphi}(\textbf{\textit{x}}). \label{Koopman_PDE}
\end{align}

According to Ref.\cite{brunton2021modern}, Koopman eigenpairs can be identified using the PDE \eqref{Koopman_PDE} based on the sparse identification of nonlinear dynamics (SINDy) framework \cite{doi:10.1073/pnas.1517384113}.
First, a library of candidate functions is chosen:
\begin{align}
    \Theta(\textbf{\textit{x}})=[\theta_1(\textbf{\textit{x}})\ \theta_2(\textbf{\textit{x}})\ ...\ \theta_L(\textbf{\textit{x}})]^{\text{T}} \label{eq_Koopman_eigenpairs}.
\end{align}

Note that $\Theta$ is often large so that Koopman eigenfunctions may be well approximated in this library:
\begin{align}
    \boldsymbol{\varphi}(\textbf{\textit{x}})\approx\boldsymbol{\Xi}\Theta(\textbf{\textit{x}}), \label{Koopman_eigenfunction_approximated_in_library}
\end{align}
where $\boldsymbol{\Xi}\in\mathbb{C}^{P\times L}$.

Given $M$ snapshots of the state $\textbf{\textit{X}}=[\textbf{\textit{x}}_{1},\textbf{\textit{x}}_{2},...,\textbf{\textit{x}}_{M}]\in\mathbb{R}^{n\times m}$ with the autonomous system dynamics Eq.\eqref{system_autonomous_dynamics}, learning Koopman eigenpairs becomes finding an optimum solution for the optimization problem as given in 
\begin{subequations}
\label{optimization}
    \begin{align}
      & \underset{\boldsymbol{\Lambda} ,\boldsymbol{\Xi} }{\mathop{\min }}\ \ \ \sum\limits_{m}{\left\| {{e}^{\boldsymbol{\Lambda} t_m}}\boldsymbol{\varphi} ({{\boldsymbol{x}}_{\text{1}}})-\boldsymbol{\varphi} ({{\boldsymbol{x}}_{m}}) \right\|}+\alpha \left\| \boldsymbol{\Xi}  \right\| \label{optimization_1}\\ 
     & s.t.\text{\ \ \ \ }\boldsymbol{\varphi}(\textbf{\textit{x}})=\boldsymbol{\Xi}\Theta(\textbf{\textit{x}}) \label{optimization_2}
    \end{align}
\end{subequations}
where $\alpha$ is a thresholding parameter to balance between sparsity and prediction accuracy.

One of the leading algorithms to solve the problem is KRONIC \cite{2017Data}, \cite{Kaiser_2021}. It proposed to identify each eigenpair separately based on the implicit formulation in Eq.\eqref{Koopman_PDE_new}.
For the $s$th eigenvalue $\lambda_s(s=1,2,...,S)$, the Koopman PDE in Eq.\eqref{Koopman_PDE} yields
\begin{align}
    {{\boldsymbol{\xi}}_{s}}({{\lambda }_{s}}\Theta (\textbf{\textit{X}})-\mathcal{T}(\textbf{\textit{X}}))=0, \label{Koopman_PDE_new}
\end{align}
where ${\boldsymbol{\xi}}_{s}$ represents the $s$th row of $\boldsymbol{\Xi}$,
\begin{align}
    \mathcal{T}(\textbf{\textit{X}})=[\nabla {{\theta }_{1}}(\textbf{\textit{X}})\cdot \dot{\textbf{\textit{x}}}_1,\nabla {{\theta}_{2}}(\textbf{\textit{X}})\cdot \dot{\textbf{\textit{x}}}_2,\cdots,\nabla {{\theta }_{L}}(\textbf{\textit{X}})\cdot \dot{\textbf{\textit{x}}}_M]^{\text{T}}
\end{align}
where $\mathcal{T}(\textbf{\textit{X}})\in\mathbb{C}^{L\times M}$. The time derivative $\dot{\textbf{\textit{X}}}=[\dot{\textbf{\textit{x}}}_{1},\dot{\textbf{\textit{x}}}_{2},...,\dot{\textbf{\textit{x}}}_{M}]$ can be measured or approximated numerically by the total variation derivative \cite{8308641}.

KRONIC proposed to identify each eigenpair separately based on the implicit formulation in Eq.\eqref{Koopman_PDE_new}. The algorithm starts with an initial guess of the eigenvalues as $\boldsymbol{\Lambda}^{init} =\mathcal{T}(\textbf{\textit{X}})\Theta^{\dagger}{(\boldsymbol{X})}$. For each ${\lambda }_{s}$, the technique subsequently alternates between an searching for the sparsest vector in the null-space of ${{\lambda }_{s}}\Theta (\boldsymbol{X})-\mathcal{T}(\textbf{\textit{X}})$ and updating of the eigenvalue ${\lambda }_{s}$ as $({\boldsymbol{\xi}_s}\mathcal{T}(\textbf{\textit{X}})\Theta^{\dagger} {(\boldsymbol{X})}\boldsymbol{\xi}^{\text{T}}_s)/(\boldsymbol{\xi }_s{\boldsymbol{\xi}^{\text{T}}_s})$, where the superscript $\dagger$ denotes the pseudoinverse operator. When $\boldsymbol{\Lambda}=diag(\lambda_1,\lambda_2,...,\lambda_S)$ converges, $\boldsymbol{\Lambda}$ and $\boldsymbol{\varphi}(\textbf{\textit{x}})=(\varphi_1,\varphi_2,...,\varphi_S)^{\text{T}}$ are identified.

\subsection{Identifying Koopman Eigenfunctions for Frequency dynamics in MIDC Systems} \label{subsec_Identifying_Koopman_Eigenfunctions}
In the following we formulate a framework to identify Koopman eigenpairs for frequency dynamics in MIDC systems directly, which unifies and extends innovations of the KRONIC algorithm by leveraging prior knowledge of power system dynamics and the idea of ensemble learning.
\subsubsection{Library Construction} \label{Integrating_nonlinearities_of_the_MIDC_system}
Selecting a proper library $\Theta$ is fundamental for identifying underlying eigenfunctions, while the wrong library functions can obscure the simplest model \cite{champion2019data}. However, for MIDC systems, selecting the best library functions is an open problem. 

Our strategy is to start with knowledge of the dynamics of MIDC systems, and to increase the complexity of the library by including more possible terms to compensate for inadequacies in modelling of system dynamics.

A general form of MIDC systems considered in this paper are the same as that in Ref.\cite{2021Optimal}. Considering the second-order dynamic models of generators and the first-order inertia models of a LCC-HVDC system \cite{Changhong2019Distributed,2020Distributed,2021Optimal}, the dynamics of MIDC systems can be written as
\begin{align}
    &\dot{\delta_i}=\omega_i, i\in \mathcal{N}_\mathcal{G}\cup\mathcal{N}_\mathcal{D} \label{eq_theta} \\
    &M_{i}\dot{\omega_i}+D_{i}\omega_i=P_{i}-\sum_{j\in \mathcal{N}}B_{ij}\sin(\delta_{i}-\delta_{j})\\
    &-\bar{k}_i^\mathcal{G}\omega_{i}, i\in \mathcal{N}_\mathcal{G}\\
    &0=P_i+p_i^{dc}-\sum_{j\in \mathcal{N}}B_{ij}\sin(\delta_i-\delta_j), i\in\mathcal{N}_\mathcal{D}\\
    &0=P_i-\sum_{j\in \mathcal{N}}B_{ij}\sin(\delta_i-\delta_j), i\in\mathcal{N}_\mathcal{P}\\
    &T_i^\mathcal{D}\dot{p}_i^{dc}={p}_i^{dc}+P_i^\mathcal{D}+u_i-k_i^\mathcal{D}\omega_i, i\in\mathcal{N}_\mathcal{D} \label{eq_droop_control_equation}
\end{align}
where three types of buses, i.e.,
generator buses, LCC-HVDC connected buses and passive load
buses are denoted by $\mathcal{N}_\mathcal{G}$, $\mathcal{N}_\mathcal{D}$ and $\mathcal{N}_\mathcal{P}$, respectively, $\mathcal{N}=\mathcal{N}_\mathcal{G}\cup\mathcal{N}_\mathcal{D}\cup\mathcal{N}_\mathcal{P}$, $\delta_i$ is the phase angle at bus $i$ with reference to the synchronous rotation coordinate, $\omega_i$ is the frequency deviation
from the nominal frequency, $M_i$ is the inertia constant of the generator $i$, $D_i>0$ is the damping coefficient, $P_i$ is the power injection ($>0$) or demand ($<0$), $p_i^{dc}$ is the transmission power of LCC-HVDC, $P^\mathcal{D}$ is the nominal value of $p^{dc}$, $u_i$ is DC power reference regulation amount when LCC-HVDC $i$ provides the frequency support for the MIDC system, $T^\mathcal{D}$ is the inertia time constant of LCC-HVDC $i$, $B_{ij}=\hat{B}_{ij}V_iV_j$ is the effective susceptance of
line $(i,j)$, $V_i$ is the voltage amplitude at bus $i$ which is assumed to be constant due to its irrelevance to the frequency control, $\bar{k}_i^\mathcal{G}>0$ is the droop coefficient of the generator $i$. When identifying eigenpairs for frequency dynamics in MIDC systems, we assume full access to the state $\boldsymbol{x}=\{\delta_h,\omega_i,{p}_i^{dc}\ |\ h\in \mathcal{N}_\mathcal{G}\cup\mathcal{N}_\mathcal{D}, i\in\mathcal{N}_\mathcal{D}\}$.

Considering system dynamics given in Eq.\eqref{eq_theta}-Eq.\eqref{eq_droop_control_equation}, nonlinearities are maily introduced by sinusoidal terms. Hence, trigonometric terms of $\boldsymbol{x}$ are included in $\Theta$ as basis functions to capture the intrinsic nonlinearities of the MIDC system. Moreover, trigonometric transform of subtraction between any two angles $\delta_i-\delta_j\  (i,j\in\mathcal{N}_\mathcal{G}\cup\mathcal{N}_\mathcal{D})$ are also included. 

Furthermore, to compensate for nonlinearities ignored in state modelling, such as deadzone setting, more possible terms should be included in $\Theta$. Here, polynomials are considered, since they represent Taylor series approximations for a broad class of smooth functions.

\subsubsection{Library Subsampling and Ensemble Learning for Eigenpairs}
\label{Library_Subsampling_and_Ensemble_Learning_for_Eigenpairs}
While KRONIC has been demonstrated on a number of examples, it faces the following problems when applied to frequency dynamics of MIDC systems.
\begin{enumerate}
    \item To make sure that the Koopman eigenfunction can be well approximated, $\Theta$ is often chosen large enough. However, in each iteration of $\boldsymbol{\Lambda}$ and $\boldsymbol{\varphi}(\textbf{\textit{x}})$, a least square solution of $\boldsymbol{\Lambda}$ should be calculated, of which the calculation complexity is $\mathcal{O}(m{{L}^{2}})$. Therefore, a large $\Theta$ is computational unfriendly.
    \item 
    Many eigenfunctions are spurious even when $\boldsymbol{\Lambda}$ converges, i.e. these eigenfunctions do not behave linearly as predicted by their corresponding eigenvalues. Therefore, verified eigenpairs should be further selected from the $S$ identified eigenpairs. A verified Koopman eigenpair $(\lambda_s,\varphi_s)$ is obtained when the evolution of the eigenfunction $\varphi_s$ on a trajectory $\textbf{\textit{X}}$ corresponds to the linear prediction using the eigenvalue $\lambda_s$, i.e., $e^{\lambda_s t_m}\varphi_{s}(\textbf{\textit{x}}_1)$. Denote the number of verified eigenpairs as $S'$. Although the number of identified eigenpairs could be very large, the number of verified eigenpairs $S'$ may still be small. A small set of verified eigenpairs with $S'<N$ is less likely to model the high dimensional nonlinear dynamics (this will be demonstrated in Section \ref{case_study}). 
\end{enumerate}

To alleviate the above problems, a library bagging method is leveraged in our Koopman eigenpairs construction method. 

Given the set of library of basis functions, it is possible to sample them to produce several different subsets $\Theta_d(d=1,2,...,D)$ and then apply identification of eigenpairs for each subset.

After acquiring $\boldsymbol{\Lambda}_d=diag(\lambda_d^1,\lambda_d^2,...,\lambda_d^{\smaller{|{\Theta}_{d}|}})$ and $\boldsymbol{\varphi}_d(\boldsymbol{x})=(\varphi_d^1,\varphi_d^2,...,\varphi_d^{\smaller{|{\Theta}_{d}|}})^\text{T}$ for each sub-library $\Theta_d$, where $\left| {{\Theta }_{d}} \right|$ represents the size of ${\Theta }_{d}$, the ensembled candidate set of eigenvalues and eigenfunctions are formed by $\boldsymbol{\Lambda}=diag(\boldsymbol{\Lambda}_1,\boldsymbol{\Lambda}_2,...,\boldsymbol{\Lambda}_D)$ and $\boldsymbol{\varphi}=[\boldsymbol{\varphi}_1,\boldsymbol{\varphi}_2,...,\boldsymbol{\varphi}_D]$. 

The prediction error defined in Eq.\eqref{verified_eigenpairs} is computed on a tested trajectory $\textbf{\textit{X}}^{\#}(t)$ to distinguish accurate eigenpairs.   
\begin{align}
    Er = \left|\varphi(\textbf{\textit{x}}^{\#}(t))-e^{\lambda t}\varphi(\textbf{\textit{x}}^{\#}(0))\right|/\left|\varphi(\textbf{\textit{x}}^{\#}(t))\right| \label{verified_eigenpairs}
\end{align}

Subsequently, identified eigenpairs can be ranked according to the error $Er$. All eigenpairs with the error below a threshold may then be used to construct the dynamic model in eigenfunction coordinates.

\vspace{0.2cm}
\textit{\textbf{Remark 1}}: Various random subsampling approaches can be used to produce sub-libraries. In our subsampling method, we first classify basis functions in the library by five categories, namely polynomials, sinusoids and  cosinusoids terms of $\textit{\textbf{x}}$, sinusoids and cosinoids of $\{\delta_i-\delta_j| i,j\in\mathcal{N}\}$. Polynomials are included in all subsets. For other four categories, each of them has a 50\% probability to be sampled out to form the subset in each sampling. 
Then we have $2^{L-1}$ subsets in total. The training of eigenpairs on different subsets is in parallel.

\vspace{0.2cm}
\textit{\textbf{Remark 2}}: Due to the different basis functions included, the acquired eigenfunctions are potential to be quite different. Specifically, introducing the library bagging technique is to execute the KRONIC framework with different initial points and search directions. Therefore, the optimization problem given in Eq.\eqref{optimization} is potential to converge to different optimum solutions with different sub-libraries, leading to an increase in the number of accurately identified eigenpairs.

The advantages to introduce library bagging technique are as follows.
\begin{enumerate}
    \item Smaller libraries can drastically speed up model identification, as the complexity of algorithm for each subset drops to $\mathcal{O}(m{{\left| {{\Theta }_{d}} \right|}^{2}})$. Library bagging can therefore help counteract the increasing computational cost of solving multiple regression problems in the ensemble. 
    \item A large set of verified eigenpairs with $S'<N$ is more likely to model the high dimensional nonlinear dynamics (this will be demonstrated in Section \ref{case_study}).
\end{enumerate}

Up to this point, the Koopman eigenpairs construction method for frequency dynamics in MIDC systems is developed.

\subsection{Koopman-Operator-Based Emergency Frequency Control Strategy}
\label{Koopman_Operator_Based_Emergency_Frequency_Control_Strategy}
In this subsection, we consider a control-affine system as
\begin{align}
    \frac{d}{dt}\textbf{\textit{x}}(t)=\textbf{\textit{f}}(\textbf{\textit{x}})+\boldsymbol{B}\boldsymbol{u}, \label{eq_control_affine_system}
\end{align}
where $\boldsymbol{u}\in\mathbb{R}^q$ is the multi-channel control input vector with $u_i(i\in\mathcal{N}_\mathcal{D})$ as entries, $\boldsymbol{B}\in\mathbb{R}^{n\times q}$ is the control matrix.

In case that the control matrix $\boldsymbol{B}$ is unknown, we first propose a method to identify it. Next, we derive how the control input affects the dynamics of these eigenfunction coordinates. Followed by this, the optimal control problem is formulated in these coordinates and a corresponding Koopman eigenfunctions feedback controller is developed. The optimal control problem yields a nonlinear control law in the original state variables. Finally, the control law is revised to adapt to local measurements.
\subsubsection{Control in eigenfunction coordinates}
We use $\boldsymbol{\Lambda}_E$ and $\boldsymbol{\varphi}_{E}$ to denote verified eigenvalues and eigenfunctions with prediction errors below $E$. According to Ref.\cite{brunton2021modern,Kaiser_2021}, with the approximated Koopman operator for an autonomous system, the control terms in Eq.\eqref{eq_control_affine_system} affect the dynamics of Koopman eigenfunctions as Eq.\eqref{eq:actual_nonlinear_system}.
\begin{align}
    \frac{d}{dt}\boldsymbol{\varphi}_{E}(\textbf{\textit{x}})
    & = \boldsymbol{\Lambda}_{E}{\boldsymbol{\varphi}}_{E}(\textbf{\textit{x}})+\nabla\boldsymbol{\varphi}_{E}(\textbf{\textit{x}})\cdot\boldsymbol{B}\boldsymbol{u} \notag \\
    & = \boldsymbol{\Lambda}_{E}{\boldsymbol{\varphi}}_{E}(\textbf{\textit{x}})+\boldsymbol{M(\textbf{\textit{x}})}\boldsymbol{u} \label{eq:actual_nonlinear_system}
\end{align}
with $\boldsymbol{M(\textbf{\textit{x}})}:=\nabla{\boldsymbol{\varphi}}_{E}(\textbf{\textit{x}})\cdot\boldsymbol{B}$.

\subsubsection{Discovering control matrix from data}
In DC power reference regulation problem for EFC, even if $\boldsymbol{B}$ can be expressed explicitly with the diagonal element corresponding to $p_i^{dc}$ as $1/T_i^\mathcal{D}$, $\boldsymbol{B}$ may still be unknown since $T_i^\mathcal{D}$ of LCC-HVDC $i$ is hard to obtain
due to dependency on operating conditions and parameter
uncertainty \cite{huang2021decentralized}. So it is of interest to discover it from data. Based on the identification of Koopman eigenpairs in Sec.\ref{subsec_Identifying_Koopman_Eigenfunctions}, $\boldsymbol{B}$ can be estimated from Eq.\eqref{Calculate_B} with sampled pairs $\{\textbf{\textit{x}}_m^u, \textbf{\textit{u}}_m\}^M _{m=1}$, where $\textbf{\textit{u}}_m$ are random control inputs which can be zero-mean white noise signals, e.g. a truncated Gaussian distribution \cite{PICALLO2022108405}.
Define
\begin{align}
    \mathcal{T}(\textbf{\textit{x}}_m^u)=[\nabla {{\theta }_{1}}({{\textbf{\textit{x}}_m^u}^{\text{T}}})\cdot \dot{\textbf{\textit{x}}}_1^u,\nabla {{\theta}_{2}}({{\textbf{\textit{x}}_m^u}^{\text{T}}})\cdot \dot{\textbf{\textit{x}}}_2^u,\cdots,\nabla {{\theta }_{p}}({{\textbf{\textit{x}}_m^u}^{\text{T}}})\cdot \dot{\textbf{\textit{x}}}_m^u] \notag
\end{align}
for $m=1,2,...,M$.
Then by rearranging Eq.\eqref{eq:actual_nonlinear_system} in terms of the library of basis functions, the control matrix $\boldsymbol{B}$ can be calculated by
\begin{equation}
  \left[
  \begin{gathered}
  (\boldsymbol{\Xi}\nabla\Theta(\textbf{\textit{x}}_1^u))\otimes\textbf{\textit{u}}_1^\text{T}\\
  (\boldsymbol{\Xi}\nabla\Theta(\textbf{\textit{x}}_2^u))\otimes\textbf{\textit{u}}_2^\text{T}\\
  \vdots\\
  (\boldsymbol{\Xi}\nabla\Theta(\textbf{\textit{x}}_M^u))\otimes\textbf{\textit{u}}_M^\text{T}
  \end{gathered}
  \right]\textbf{\textit{b}}=\left[
  \begin{gathered}
  \boldsymbol{\Xi}\mathcal{T}(\textbf{\textit{x}}_1^u)\\
  \boldsymbol{\Xi}\mathcal{T}(\textbf{\textit{x}}_2^u)\\
  \vdots\\
  \boldsymbol{\Xi}\mathcal{T}(\textbf{\textit{x}}_M^u)
  \end{gathered}
  \right]-\left[
  \begin{gathered}
  \boldsymbol{\Lambda}\boldsymbol{\Xi}\Theta(\textbf{\textit{x}}_1^u))\\
  \boldsymbol{\Lambda}\boldsymbol{\Xi}\Theta(\textbf{\textit{x}}_2^u))\\
  \vdots\\
  \boldsymbol{\Lambda}\boldsymbol{\Xi}\Theta(\textbf{\textit{x}}_M^u))
  \end{gathered} \label{Calculate_B}
  \right]
\end{equation}
with $\textbf{\textit{b}}=[\textbf{\textit{b}}_1,...,\textbf{\textit{b}}_n]^\text{T}$
where $\textbf{\textit{b}}_n$ represents the $n$th row of $\boldsymbol{B}$ and $\otimes$ is the Kronecker product. Here, the subscript $E$ is omitted for $\boldsymbol{\Lambda}$ and $\boldsymbol{\Xi}$, where $\boldsymbol{\Lambda}_{E}$ is defined in Eq.\eqref{eq:actual_nonlinear_system} and $\boldsymbol{\Xi}_{E}$ represents the coefficient matrix for $\boldsymbol{\varphi}_{E}$. While $\boldsymbol{\Xi}$ and $\boldsymbol{\Lambda}$ have been discovered in Section \ref{subsec_Identifying_Koopman_Eigenfunctions}, $\Theta$ and $\mathcal{T}$ have been evaluated on the sampled pairs $\{\textbf{\textit{x}}_m^u, \textbf{\textit{u}}_m\}^M _{m=1}$. Therefore, a least-squares solution of $\textbf{\textit{b}}$ can be calculated by Eq.\eqref{Calculate_B}.

\vspace{0.2cm}
\textit{\textbf{Remark 1}}: The random control input can also be other kinds of distribution, e.g. a uniform distribution.

\subsubsection{Formulation of the optimal control problem}
We now design the LQR controller based on verified Koopman eigenfunctions and the identified control matrix. For brevity, the subscript $E$ are omitted later in this section. The control objective is a quadratic cost functional:
\begin{gather}
    J(\boldsymbol{\varphi},\textbf{\textit{u}}) = \int_0^\infty \boldsymbol{\varphi}^\text{T}(\textbf{\textit{x}})\boldsymbol{Q}\boldsymbol{\varphi}(\textbf{\textit{x}})+\boldsymbol{u}^\text{T}\boldsymbol{R}\boldsymbol{u}\,dt.  \label{cost_function}
\end{gather}
where the structure of $\boldsymbol{Q}$ is chosen such that it only minimizes the norm of frequency. If $\boldsymbol{x}=\{\delta_h,\omega_i,{p}_i^{dc}\ |\ h\in \mathcal{N}_\mathcal{G}\cup\mathcal{N}_\mathcal{D}, i\in\mathcal{N}_\mathcal{D}\}$ are locally measurable when regulating power reference for LCC-HVDC $i$, EKEFC can be formulated as an optimization problem with Eq.\eqref{cost_function} as the control objective and Eq.\eqref{eq:actual_nonlinear_system} as constraints.

\subsubsection{Partial measurements}
\label{Partial_measurements}
Note that the full access to $\boldsymbol{x}=\{\delta_h,\omega_i,{p}_i^{dc}\ |\ h\in \mathcal{N}_\mathcal{G}\cup\mathcal{N}_\mathcal{D}, i\in\mathcal{N}_\mathcal{D}\}$ requires wide area measurements, which is usually not an option due to the cost prohibitive communication infrastructure requirements. Therefore, it’s realistic to design distributed control with partial measurements, in which only system frequency $\omega_i$ and $p_i^{dc}\ (i\in \mathcal{N}_\mathcal{D})$ are known. Define $\boldsymbol{X}'_i=\{x'_i|x'_i=(\omega_s,p_i^{dc}), s\in\mathcal{N}_\mathcal{D}\}$ for LCC-HVDC $i$. A specific set of eigenfunctions for LCC-HVDC $i$ can be selected, in which only eigenfunctions explicitly expressed in terms of $\boldsymbol{X}'_i$ are included. We use ${\boldsymbol{\Lambda }}_i$ and ${\boldsymbol{\varphi}}_i$ to denote the specific set of eigenvalues and the corresponding set of eigenfunctions for LCC-HVDC $i$. We make the following assumption.

\vspace{0.2cm}
\textit{\textbf{Assumption 1}}: The frequency dynamics in any node or generator are the same.
Namely, spatio-temporal distribution characteristics of frequency dynamics can be neglected.

Based on the above assumption, if only $\omega_i$ and $p_i^{dc}$ is measurable at LCC-HVDC $i$, $\omega_s\ (s\in\mathcal{N}_\mathcal{D})$ in ${\boldsymbol{\varphi}}_i$ take the value of $\omega_i$ for the optimal control strategy calculation for LCC-HVDC $i$. Even though $\textit{\textbf{Assumption 1}}$ may weaken the control effect of EKEFC, the robustness of EKEFC and the upper bound for error in the eigenvalues of controlled system dynamics, which will be discussed in the next section, may alleviate this problem.

Up to this point, EKEFC of LCC-HVDC $i$ to the regulate DC power reference can be expressed as
\begin{align}
  \min \text{ }\int_{0}^{\infty }&{{{({{\boldsymbol{\varphi }}_{i}}(\boldsymbol{X}'_i)-{\boldsymbol{\varphi }}_{i,ref})}^{\text{T}}}\boldsymbol{Q}_i}({{\boldsymbol{\varphi }}_{i}}(\boldsymbol{X}'_i)-{{\boldsymbol{\varphi }}_{i,ref}})\notag\\
  &\boldsymbol{+}u_{i}^{\text{T}}R{{u}_{i}}\text{ }dt \notag
\end{align}
\begin{align}
 & \text{s.t.}\text{  }\frac{d{{\boldsymbol{\varphi }}_{i}}(\boldsymbol{X}'_i)}{dt}\boldsymbol{=}{{\boldsymbol{\Lambda }}_{i}}{{\boldsymbol{\varphi }}_{i}}(\boldsymbol{X}'_i)+\boldsymbol{M}_i(\boldsymbol{X}'_i){u}_{i} \notag\\
 &\ \ \ \ \ \ \boldsymbol{M}_i(\boldsymbol{X}'_i)=\nabla{\boldsymbol{\varphi}}_i(\boldsymbol{X}'_i)\cdot\boldsymbol{B}_i
 \label{Koopman_based_LQR_for_HVDC}
\end{align}
where $\boldsymbol{\varphi}_{i,ref}$ is the reference of $\boldsymbol{\varphi}$ when the frequency reaches its nominal value, $\boldsymbol{Q}_i$ is a diagonal weight matrix for koopman eigenfunctions. It has non-zero diagonal entries when the corresponding eigenfunction only explicitly expressed in terms of frequencies, $\boldsymbol{B}_i$ is the column in $\boldsymbol{B}$ corresponding to LCC-HVDC $i$.
The optimum of Eq.\eqref{Koopman_based_LQR_for_HVDC} can be solved by a state-dependent Ricatti equation (SDRE) given as 
\begin{align}
    u_{i}^{*}=-{{R}^{-1}}\boldsymbol{M}_i^{\text{T}}(\boldsymbol{X}'_i)\boldsymbol{H}{{\boldsymbol{\varphi }}_{i}}(\boldsymbol{X}'_i), \label{Koopman_based_SDRE1}
\end{align}
where $u_i$ is DC power reference regulation amount when LCC-HVDC $i$ provides the frequency support for the MIDC system, $\boldsymbol{H}$ satisfies
\begin{align}
    \boldsymbol{Q}+\boldsymbol{H}{{\boldsymbol{\Lambda }}_{i}}+{\boldsymbol{\Lambda }}_{i}\boldsymbol{H}-\boldsymbol{H}\boldsymbol{M}_i^{\text{T}}(\boldsymbol{X}'_i)R^{-1}\boldsymbol{M}_i^{\text{T}}(\boldsymbol{X}'_i)\boldsymbol{H}=0. \label{Koopman_based_SDRE}
\end{align}

The EKEFC strategies can be applied online as follows. LCC-HVDC $i$ measures $x_i'=(\omega_i,p_i^{dc})$ at bus $i\ (i\in\mathcal{N}_\mathcal{D})$ periodically. When an emergency is detected and EKEFC is enabled at $t_{1}$, the control input can be determined by solving Eq.\eqref{Koopman_based_SDRE1} and Eq.\eqref{Koopman_based_SDRE} online every $\Delta t_1$ over which the applied control is kept constant. Denote the computation time of EKEFC as $\Delta t_2$. After $x_i'$ is measured at $t_{k}(k=1,2,...)$, the optimal active power reference can be obtained at $t_{k}+\Delta t_{2}$. The optimal active power reference is then applied to LCC-HVDC $i$ during $[t_{k}+\Delta t_{2}, t_{k+1}+\Delta t_{2}]$, where $t_{k+1}=t_{k}+\Delta t_1$. Note that $\Delta t_{1}$ should be set strictly larger than $\Delta t_{2}$.

\begin{figure*}[h] 
    \centering
    \includegraphics[width=15cm]{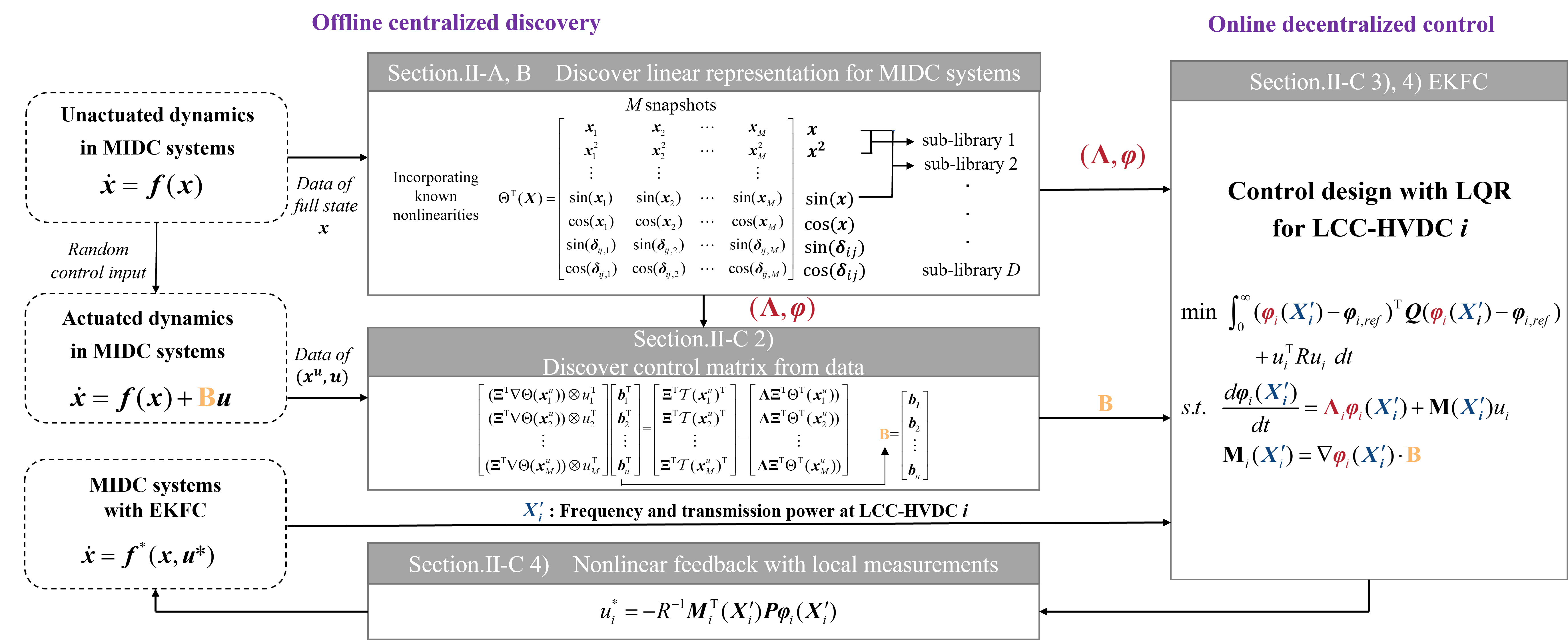}
    \caption{Frequency Control Framework of MIDC systems via Koopman linear representation}\label{fig_Frequency_Control_Framework_of_MIDC_system_via_Koopman_linear_representation}
\end{figure*}

In general, our data-driven EKEFC framework is shown in Fig.\ref{fig_Frequency_Control_Framework_of_MIDC_system_via_Koopman_linear_representation}.

\section{Robustness Analysis and Error Estimation} \label{Robustness_Analysis_and_Error_Estimation}
In this section, we first establish the robustness properties of the proposed KEFC. Next, we examine the effect of an error in the representation of the Koopman operator on the close-loop dynamics and provide an error estimation method. For brevity, the subscript $E$, which is the threshold for prediction errors, and the subscript $i$, which represents LCC-HVDC $i$ will be omitted in this section.

In Section \ref{Emergency_Frequency_Controller_Design}, the formulation of optimal control problem Eq.\eqref{Koopman_based_LQR_for_HVDC}-Eq.\eqref{Koopman_based_SDRE} in coordinates of eigenfunctions is based on the assumption that Koopman eigenpairs $(\boldsymbol{\Lambda},\boldsymbol{\varphi}(\textbf{\textit{x}}))$ are accurate. However, in real applications, there are inevitable representation errors in both $\boldsymbol{\Lambda}$ and $\boldsymbol{\varphi}(\textbf{\textit{x}})$, leading to gap between $\dot{\boldsymbol{\varphi}}$ and $\boldsymbol{\Lambda}\boldsymbol{\varphi}(\textbf{\textit{x}})$. There are three possible sources of the representation errors \cite{guo2015error}. For the representation of eigenpairs for dynamics in MIDC systems, the three possible sources of the representation errors are listed as follows.
\begin{enumerate}
    \item The first source of representation errors is the reconstruction error of eigenfunctions. In our method, candidate functions in library are selected by integrating prior knowledge in dynamics of generators and LCC-HVDC system. However, we can hardly guarantee that the library is rich enough to include all categories of nonlinearities required for construction of eigenfuntions.
    \item The second source of representation errors is the training error. In EKEFC, the training of eigenvalues terminates when the training error is less than a specified tolerance. The training of eigenfunctions for each eigenvalue is to select an eigenfuntion with minimum prediction error. Therefore, inadequate training leads to representation errors in Koopman eigenpairs. 
    \item The third source of representation errors is from sampling. The exact approximation of either eigenpairs or the control matrix requires that the approximation to be based on the entire state space and the entire control space. However, the approximation is based on sampling. Specifically, in practical online operational contexts in EKEFC, there are numerous unpredicted conditions, such as uncertain emergency faults and time-delay measurements. However, it's impossible for collected datasets to cover all the conditions above. Hence, there is an unavoidable bias error due to incomplete sampling of the state space. 
\end{enumerate}

These inevitable representation errors not only affect representation accuracy for eigenpairs, but impact the control effect of the feedback controller. Specifically, there are two questions we need to clarify when taking consideration of representation errors:
\begin{enumerate}
    \item First, is the controlled system robust to representation errors in eigenpairs? In another word, would an arbitrarily small error destablize the close-loop system?
    \item  Second, if the closed-loop system is robust to representation errors in eigenpairs, how can we guarantee the dynamics of the controlled system under misrepresented Koopman operator is closed to the ones under the accurate Koopman operartor?
\end{enumerate}

Before answering the two questions, we need to figure out how to model representation errors of the Koopman eigenpairs. According to the three sources of representation errors, the gap between $\dot{\boldsymbol{\varphi}}$ and $\boldsymbol{\Lambda}\boldsymbol{\varphi}(\textbf{\textit{x}})$ must be taken into consideration, as well as misrepresentation of both eigenvalues and eigenfunctions.

Assume eigenvalues and eigenfunctions with representation errors can be expressed as 
\begin{gather}
    \hat{\boldsymbol{\Lambda}}:=\boldsymbol{\Lambda}+\boldsymbol{\varepsilon_{\boldsymbol{\Lambda}}} \label{eq:lambda_hat}\\
    \hat{\boldsymbol{\varphi}}(\textbf{\textit{x}}):=\boldsymbol{\varphi}(\textbf{\textit{x}})+\boldsymbol{\varepsilon}_{\boldsymbol{\varphi}}\boldsymbol{\psi}(\textbf{\textit{x}}) \label{eq:phi_hat}
\end{gather}
where $\boldsymbol{\varepsilon}_{\boldsymbol{\Lambda}}\in\mathbb{C}^{n\times n}$ is a diagonal matrix representing the discrepancy to the true $\boldsymbol{\Lambda}$; $\boldsymbol{\varepsilon}_{\boldsymbol{\varphi}}\boldsymbol{\psi}(\textbf{\textit{x}})$ is the discrepancy to the eigenfunctions ${\boldsymbol{\varphi}}(\textbf{\textit{x}})$, where $\boldsymbol{\varepsilon}_{\boldsymbol{\varphi}}$ is a matrix with very small numbers as diagonal entries and ${\boldsymbol{\varphi}}(\textbf{\textit{x}})\neq\boldsymbol{\psi}(\textbf{\textit{x}})$.


\vspace{0.2cm}
\textit{\textbf{Remark 1}}: Even if accurate eigenpairs are obtained, if there are errors in measurements of $\textbf{\textit{x}}$, such as time-delay measurements or noises, the control effects may still be weakened. If measurements with errors can be expressed in terms of $\textbf{\textit{x}}$, then eigenpairs can be written as $(\boldsymbol{\Lambda},\boldsymbol{\varphi}(\kappa(\textbf{\textit{x}})))$, where $\kappa(\textbf{\textit{x}})$ represents measurements with errors. In this case, $\boldsymbol{\varphi}(\kappa(\textbf{\textit{x}}))$ can be regarded as $\hat{\boldsymbol{\varphi}}(\textbf{\textit{x}})$. In Section \ref{Partial_measurements}, $\omega_s\ (s\in\mathcal{N}_\mathcal{D})$ in ${\boldsymbol{\varphi}}_i$ take the value of $\omega_i$. In this case, $\kappa(\textbf{\textit{x}})$ is a $n$-dimensional vector where each entry equals to $\mathbf{1}_i\cdot\textbf{\textit{x}}=\omega_i$, where $\mathbf{1}_i$ is a unit vector. Thus, the following analysis of misrepresented eigenpairs can also be used to analyse the effect of errors in measurements. 
\vspace{0.2cm}

We examine the effect of misrepresentation of eigenpairs with an $\boldsymbol{\beta}(\textbf{\textit{x}})$ given as
\begin{align}
     \boldsymbol{\beta}(\textbf{\textit{x}})=\dot{\hat{\boldsymbol{\varphi}}}(\textbf{\textit{x}})-\hat{\boldsymbol{\Lambda}}\hat{\boldsymbol{\varphi}}(\textbf{\textit{x}}) 
\end{align}
where $(\hat{\boldsymbol{\Lambda}},\hat{\boldsymbol{\varphi}}(\textbf{\textit{x}}))$ are eigenpairs learned from data.
Accordingly, the dynamics of a misrepresented eigenfunction satisfies
\begin{align}
     \dot{\hat{\boldsymbol{\varphi}}}(\textbf{\textit{x}})=\hat{\boldsymbol{\Lambda}}\hat{\boldsymbol{\varphi}}(\textbf{\textit{x}})+\boldsymbol{\beta}(\textbf{\textit{x}}). \label{eq:dynamics_in_error}
\end{align}

Combining Eq.(\ref{eq:dynamics_in_error}) and  Eq.(\ref{eq_autonomous_dynamics}), $\boldsymbol{\beta}(\boldsymbol{x})$ can be expressed as a function of $(\hat{\boldsymbol{\Lambda}},\hat{\boldsymbol{\varphi}}(\textbf{\textit{x}}))$ given as
\begin{align}
    \boldsymbol{\beta}(\boldsymbol{x})=\boldsymbol{\varepsilon_{\varphi}}\dot{\boldsymbol{\psi}}(\boldsymbol{x})-\boldsymbol{\Lambda}\boldsymbol{\varepsilon_{\varphi}}-\boldsymbol{\varepsilon_{\boldsymbol{\Lambda}}}-\boldsymbol{\varepsilon_{\boldsymbol{\Lambda}}}\boldsymbol{\varepsilon_{\varphi}}\boldsymbol{\psi}(\boldsymbol{x}).
\end{align}

To answer the first question, we have the following proposition to guarantee the robustness of the closed-loop system with EKEFC to errors in the representation of Koopman eigenpairs.

\vspace{0.2cm}
\noindent \textbf{Proposition 1.} Let the system dynamics in
accurate eigenfunction coordinates be given by Eq.(\ref{eq:actual_nonlinear_system}). Then, the closed-loop solution of SDRE with the system dynamics given by Eq.(\ref{eq:dynamics_in_error}) is semiglobally asymptotically stable as long as Eq.(\ref{eq_tilde_P}) holds.
\begin{gather}
    \tilde{\boldsymbol{H}}<\dot{\boldsymbol{H}}-\boldsymbol{Q}-\boldsymbol{H}\boldsymbol{M}\boldsymbol{R}^{-1}\boldsymbol{M}^{\text{T}}\boldsymbol{H} \label{eq_tilde_P}
\end{gather}
where $\boldsymbol{H}$ satisfies a SDRE given as
\begin{gather}
    \mathbf{Q}+\boldsymbol{H}\mathbf{\Lambda}+\mathbf{\Lambda}^{\text{T}}\boldsymbol{H}-\boldsymbol{H}\boldsymbol{M}\boldsymbol{R}^{-1}\boldsymbol{M}^{-1}\boldsymbol{H}=0, \label{eq:state_dependent_ricatti}
\end{gather}
and
\begin{align}
\tilde{\boldsymbol{H}}&=\boldsymbol{H}\boldsymbol{\varepsilon_{\varphi}}\boldsymbol{N}\boldsymbol{R}^{-1}\boldsymbol{N}^{\text{T}}\boldsymbol{\varepsilon_{\varphi}}^{\text{T}}\boldsymbol{H}
\end{align}
with $\boldsymbol{N(\textbf{\textit{x}})}:=\nabla{\boldsymbol{\psi}}(\textbf{\textit{x}})\cdot\boldsymbol{B(\textbf{\textit{x}})}$.

\vspace{0.2cm}
\noindent \textit{Proof}. According to Ref.\cite{1996Nonlinear}, $V=\boldsymbol{\varphi^{\text{T}}(x)H\varphi(x)}$ is a candidate Lyapunov function of the system in accurate eigenfunction coordinates, where $\boldsymbol{H}$ is the solution of the SDRE and $V>0$. For control-affine system described by accurate eigenpairs, the derivation of $\dot{V}$ is given as
\begin{align}
    \dot{V}&=\boldsymbol{\varphi}^{\text{T}}(\boldsymbol{x})\boldsymbol{\dot{H}}\varphi(\boldsymbol{x})+\boldsymbol{\varphi}^{\text{T}}(\boldsymbol{x})\boldsymbol{H}\dot{\boldsymbol{\varphi}}(\boldsymbol{x})+\dot{\boldsymbol{\varphi}}^{\text{T}}(\boldsymbol{x})\boldsymbol{H}\boldsymbol{\varphi}(\boldsymbol{x}) \notag\\
    &=\boldsymbol{\varphi}^{\text{T}}(\boldsymbol{x})(\dot{\boldsymbol{H}}-\boldsymbol{Q}-\boldsymbol{H}\boldsymbol{M}\boldsymbol{R}^{-1}\boldsymbol{M}^{\text{T}}\boldsymbol{H})\boldsymbol{\varphi}(\boldsymbol{x})<0.
\end{align}

To analysis the impact of eigenpairs representation errors on $V$, we transform the dynamics of system in misrepresented eigenfunction coordinates into the accurate ones. Specifically, applying derivative rules to Eq.(\ref{eq:dynamics_in_error}) yields (here without input $\boldsymbol{u} = \boldsymbol{0}$)
\begin{gather}
   \dot{\hat{\boldsymbol{\varphi}}}(\textbf{\textit{x}})=\boldsymbol{\Lambda}\boldsymbol{\varphi}(\textbf{\textit{x}})+\boldsymbol{\varepsilon}_{\boldsymbol{\varphi}}\nabla{\boldsymbol{\psi}}\cdot \boldsymbol{f}(\textbf{\textit{x}}) \label{eq:deriv rule_29}
\end{gather}

Combined with Eq.(\ref{eq:dynamics_in_error}), Eq.(\ref{eq:deriv rule_29}) leads to the relationship between $\boldsymbol{\varepsilon}_{\boldsymbol{\Lambda}}$, $\boldsymbol{\varepsilon}_{\boldsymbol{\varphi}}\boldsymbol{\psi}(\textbf{\textit{x}})$ and $\boldsymbol{\beta}(\textbf{\textit{x}})$
\begin{gather}
  \boldsymbol{\varepsilon}_{\boldsymbol{\varphi}}\nabla{\boldsymbol{\psi}(\textbf{\textit{x}})}\cdot f(\textbf{\textit{x}})=\boldsymbol{\beta}(\textbf{\textit{x}})-\boldsymbol{\Lambda}\boldsymbol{\varphi}(\textbf{\textit{x}})+\hat{\boldsymbol{\Lambda}}\hat{\boldsymbol{\varphi}}(\textbf{\textit{x}}) \label{eq:}
\end{gather}

If we augment the uncontrolled system Eq.(\ref{eq:dynamics_in_error}) with a linear control term $\boldsymbol{B}\boldsymbol{u}$, the dynamics of the eigenfunction in the corresponding control-affine system can be obtained as
\begin{align}
    \dot{\hat{\boldsymbol\varphi}}(\textbf{\textit{x}})&=\boldsymbol{\Lambda}\boldsymbol{\varphi}(\textbf{\textit{x}})+ \boldsymbol{M(\textbf{\textit{x}})}\boldsymbol{u}+\boldsymbol{\varepsilon}_{\boldsymbol{\varphi}}\nabla{\boldsymbol{\psi}(\textbf{\textit{x}})}(\boldsymbol{f}(\textbf{\textit{x}})+\boldsymbol{B}\boldsymbol{u})\notag \\
    &=\hat{\boldsymbol{\Lambda}}\hat{\boldsymbol{\varphi}}(\textbf{\textit{x}})+\boldsymbol{M(\textbf{\textit{x}})}\boldsymbol{u}+\boldsymbol{\beta}(\textbf{\textit{x}})+\boldsymbol{\varepsilon}_{\boldsymbol{\varphi}}\boldsymbol{N(\textbf{\textit{x}})}\boldsymbol{u} \notag \\
    &=(\hat{\boldsymbol{\Lambda}}+\frac{\boldsymbol{\beta}(\textbf{\textit{x}})}{\hat{\boldsymbol{\varphi}}})\hat{\boldsymbol{\varphi}}+(\boldsymbol{M(\textbf{\textit{x}})}+\boldsymbol{\varepsilon}_{\boldsymbol{\varphi}}\boldsymbol{N(\textbf{\textit{x}})})\boldsymbol{u}
    \label{eq:phi_hat_dynamics}
\end{align}

By applying Eq.(\ref{eq:lambda_hat}) and Eq.(\ref{eq:phi_hat}), modelled system dynamics given in Eq.(\ref{eq:phi_hat_dynamics}) can be transformed into $\boldsymbol{\varphi}$ coordinates:
\begin{align}
    \dot{\boldsymbol{\varphi}}(\textbf{\textit{x}})=\boldsymbol{\Lambda}\boldsymbol{\varphi}(\textbf{\textit{x}})+\Bar{\boldsymbol{M}}\boldsymbol{u} \label{eq:learnt_nonlinear_system}
\end{align}
where
\begin{gather}
    \Bar{\boldsymbol{M}}=\boldsymbol{M(\textbf{\textit{x}})}+\boldsymbol{\varepsilon}_{\boldsymbol{\varphi}}\boldsymbol{N(\textbf{\textit{x}})}.
\end{gather}

A misrepresentation of the Koopman eigenpairs will affect the value of $\dot{V}$. Specifically, by applying Eq.(\ref{eq:learnt_nonlinear_system}), we have
\small{
\begin{align}
    &\dot{V}=\boldsymbol{\varphi}^{\text{T}}(\boldsymbol{x})\boldsymbol{\dot{H}}\boldsymbol{\varphi}(\boldsymbol{x})+\boldsymbol{\varphi}^{\text{T}}(\boldsymbol{x})\boldsymbol{H}\dot{\boldsymbol{\varphi}}(\boldsymbol{x})+\dot{\boldsymbol{\varphi}}^{\text{T}}(\boldsymbol{x})\boldsymbol{H}\boldsymbol{\varphi}(\boldsymbol{x}) \notag\\
    &={\boldsymbol{\varphi}}^{\text{T}}(x)\boldsymbol{\dot{H}}\boldsymbol{\varphi}(\boldsymbol{x})+\boldsymbol{\varphi}^{\text{T}}(\boldsymbol{x})\boldsymbol{H}\Big[\boldsymbol{\Lambda}\boldsymbol{\varphi}(\boldsymbol{x})-\boldsymbol{M}\boldsymbol{R}^{-1}\bar{\boldsymbol{M}}^{\text{T}}\boldsymbol{H}\boldsymbol{\varphi}(\boldsymbol{x})\Big] \notag\\
    &\ \ \ +\Big[\boldsymbol{\Lambda}\boldsymbol{\varphi}(\boldsymbol{x})-\boldsymbol{M}\boldsymbol{R}^{-1}\bar{\boldsymbol{M}}^{\text{T}}\boldsymbol{H}\boldsymbol{\varphi}(\boldsymbol{x})\Big]^{\text{T}}\boldsymbol{H}\boldsymbol{\varphi}(\boldsymbol{x}) \notag\\
    &=\boldsymbol{\varphi}^{\text{T}}(\boldsymbol{x})\cdot \notag\\
    &\big(\dot{\boldsymbol{H}}-\boldsymbol{Q}-\boldsymbol{H}\boldsymbol{M}\boldsymbol{R}^{-1}\boldsymbol{M}^{\text{T}}\boldsymbol{H}+\boldsymbol{H}\boldsymbol{\varepsilon_{\varphi}}\boldsymbol{N}\boldsymbol{R}^{-1}\boldsymbol{N}^{\text{T}}\boldsymbol{\varepsilon_{\varphi}}^{\text{T}}\boldsymbol{H}\big)\boldsymbol{\varphi}(\boldsymbol{x})
\end{align}}

According to Ref.\cite{1996Nonlinear}, $\dot{\boldsymbol{H}}-\boldsymbol{Q}-\boldsymbol{H}\boldsymbol{M}\boldsymbol{R}^{-1}\boldsymbol{M}^{\text{T}}\boldsymbol{H}<\boldsymbol{0}$. Note that though $\tilde{\boldsymbol{H}}$ is semi-positive, as long as $\dot{\boldsymbol{H}}-\boldsymbol{Q}-\boldsymbol{H}\boldsymbol{M}\boldsymbol{R}^{-1}\boldsymbol{M}^{\text{T}}\boldsymbol{H}+\tilde{\boldsymbol{H}}<\boldsymbol{0}$ holds, the error in the representation of Koopman eigenpairs would not destabilize the close-loop system. Note that the eigenvalues of a matrix depend continuously on its entries. Therefore, when eigenvalues of $\dot{\boldsymbol{H}}-\boldsymbol{Q}-\boldsymbol{H}\boldsymbol{M}\boldsymbol{R}^{-1}\boldsymbol{M}^{\text{T}}\boldsymbol{H}$ are negative, eigenvalues of $\dot{\boldsymbol{H}}-\boldsymbol{Q}-\boldsymbol{H}\boldsymbol{M}\boldsymbol{R}^{-1}\boldsymbol{M}^{\text{T}}\boldsymbol{H}+\tilde{\boldsymbol{H}}$ are still negative as long as $\left\|\boldsymbol{\varepsilon_{\varphi}}\right\|$ is sufficiently small.$\hfill\blacksquare$

\vspace{0.2cm}
In order to answer the second question, we examine the effect of misrepresented Koopman eigenpairs on the close-loop dynamics and provide an estimation of the upper error bound.

For accurate eigenpairs $(\boldsymbol{\Lambda},\boldsymbol{\varphi}(\textbf{\textit{x}}))$, the feedback control is given by 
\begin{align}
    \boldsymbol{u}&=-\boldsymbol{R^{-1}M^{\text{T}}}\boldsymbol{H}\boldsymbol{\varphi}
\end{align}
and the resulting closed-loop dynamics can be obtained as
\begin{align}
    \dot{\boldsymbol\varphi}(\textbf{\textit{x}})&=\boldsymbol{\boldsymbol{\Lambda}\varphi}(\textbf{\textit{x}})+\boldsymbol{M}\boldsymbol{u} \notag \\
    &=\big(\boldsymbol{\Lambda}-\boldsymbol{M}\boldsymbol{R}^{-1}\boldsymbol{M}^{\text{T}}\boldsymbol{H}\big)\boldsymbol{\varphi}(\textbf{\textit{x}}).
\end{align}

Similarly, for eigenpairs with representation errors $(\hat{\boldsymbol{\Lambda}},\hat{\boldsymbol{\varphi}}(\textbf{\textit{x}}))$, the control input is given by
\begin{align}
    \bar{\boldsymbol{u}}&=-\boldsymbol{R^{-1}\bar{M}^{\text{T}}}\bar{\boldsymbol{H}}\boldsymbol{\varphi}
\end{align}
where $\boldsymbol{\bar{H}}$ satisfies a SDRE with $\bar{\boldsymbol{M}}$ given as
\begin{gather}
    \mathbf{Q}+\boldsymbol{\bar{H}}\mathbf{\Lambda}+\mathbf{\Lambda}^{\text{T}}\boldsymbol{\bar{H}}-\boldsymbol{\bar{H}}\bar{\boldsymbol{M}}\boldsymbol{R}^{-1}\bar{\boldsymbol{M}}^{-1}\boldsymbol{\bar{H}}=0, \label{eq:state_dependent_ricatti}
\end{gather}
and the resulting closed-loop dynamics can be obtained as
\begin{align}
    \dot{\boldsymbol\varphi}(\textbf{\textit{x}})&=\boldsymbol{\boldsymbol{\Lambda}\varphi}(\textbf{\textit{x}})+\boldsymbol{M}\boldsymbol{u} \notag \\
    &=\big(\boldsymbol{\Lambda}-\bar{\boldsymbol{M}}\boldsymbol{R}^{-1}\bar{\boldsymbol{M}}^{\text{T}}\bar{\boldsymbol{H}}\big)\boldsymbol{\varphi}(\textbf{\textit{x}}).
\end{align}

Define $\boldsymbol{\mu}:=\boldsymbol{\Lambda}-\boldsymbol{M}\boldsymbol{R}^{-1}\boldsymbol{M}^{\text{T}}\boldsymbol{H}$ and $\boldsymbol{\hat{\mu}}:=\boldsymbol{\Lambda}-\bar{\boldsymbol{M}}\boldsymbol{R}^{-1}\bar{\boldsymbol{M}}^{\text{T}}\bar{\boldsymbol{H}}$. The estimation of error in $\boldsymbol{\mu}$ can be calculated as
\begin{align}
    &\left\|\boldsymbol{\mu}-\boldsymbol{\hat{\mu}}\right\|\notag\\
    &=\left\|-\boldsymbol{M}\boldsymbol{R}^{-1}\boldsymbol{M}^{\text{T}}\boldsymbol{H}+\bar{\boldsymbol{M}}\boldsymbol{R}^{-1}\bar{\boldsymbol{M}}^{\text{T}}\bar{\boldsymbol{H}}\right\|\notag\\
    &=\left\|-\boldsymbol{M}\boldsymbol{R}^{-1}\boldsymbol{M}^{\text{T}}\boldsymbol{H}+(\boldsymbol{M}+\boldsymbol{\varepsilon_{\varphi}}\boldsymbol{N})\boldsymbol{R}^{-1}(\boldsymbol{M}+\boldsymbol{\varepsilon_{\varphi}}\boldsymbol{N})^{\text{T}}\bar{\boldsymbol{H}}\right\|\notag\\
    &=\big\|\boldsymbol{M}\boldsymbol{R}^{-1}\boldsymbol{M}^{\text{T}}(\bar{\boldsymbol{H}}-\boldsymbol{H})+\boldsymbol{\varepsilon_{\varphi}}(\boldsymbol{N}\boldsymbol{R}^{-1}\boldsymbol{M}^{\text{T}}+\boldsymbol{M}\boldsymbol{R}^{-1}\boldsymbol{N}^{\text{T}})\bar{\boldsymbol{H}} \notag\\
    &\ \ \ +\boldsymbol{\varepsilon}^2_{\boldsymbol{\varphi}}\boldsymbol{N}\boldsymbol{R}^{-1}\boldsymbol{N}^{\text{T}}\bar{\boldsymbol{H}}\big\|\notag\\
    &\le\left\|\boldsymbol{M}\boldsymbol{R}^{-1}\boldsymbol{M}^{\text{T}}\right\|\left\|\bar{\boldsymbol{H}}-\boldsymbol{H}\right\|\notag\\
    &\ \ \ +\left\|\boldsymbol{\varepsilon_{\varphi}}\right\|\Big(\left\|\boldsymbol{N}\boldsymbol{R}^{-1}\boldsymbol{M}^{\text{T}}\bar{\boldsymbol{H}}\right\|+\left\|\boldsymbol{M}\boldsymbol{R}^{-1}\boldsymbol{M}^{\text{T}}\bar{\boldsymbol{H}}\right\|\Big)\notag\\
    &\ \ \ +\left\|\boldsymbol{\varepsilon_{\varphi}}\right\|^{2}\left\|\boldsymbol{N}\boldsymbol{R}^{-1}\boldsymbol{N}^{\text{T}}\bar{\boldsymbol{H}}\right\|\notag\\
 \label{eq_error_bound_in_mu}
\end{align}
where $\|\cdot\|$ represents the Frobenius norm. 

The upper bound of $\left\|\bar{\boldsymbol{H}}-\boldsymbol{H}\right\|$ on the right hand side is remained to be estimated explicitly in terms of representation errors of Koopman eigenpairs. Let $(\boldsymbol{\Lambda},\boldsymbol{Q},\bar{\boldsymbol{M}}\boldsymbol{R}^{-1}\bar{\boldsymbol{M}}^{\text{T}})$ be a triplet of
so-called perturbed system matrices. When the continuity condition on the sequence of $(\boldsymbol{\Lambda},\boldsymbol{Q},\boldsymbol{M}\boldsymbol{R}^{-1}\boldsymbol{M}^{\text{T}})_{\pi\rightarrow\pi_{0}}\rightarrow(\boldsymbol{\Lambda},\boldsymbol{Q},\bar{\boldsymbol{M}}\boldsymbol{R}^{-1}\bar{\boldsymbol{M}}^{\text{T}})$ holds, where $\pi\rightarrow\pi_{0}$ represents a mapping representing a deformation process from $(\boldsymbol{\Lambda},\boldsymbol{Q},\boldsymbol{M}\boldsymbol{R}^{-1}\boldsymbol{M}^{\text{T}})$ to $(\boldsymbol{\Lambda},\boldsymbol{Q},\bar{\boldsymbol{M}}\boldsymbol{R}^{-1}\bar{\boldsymbol{M}}^{\text{T}})$. By Theorem 2.5 in Ref.\cite{bishop2018robustness}, the upper bound of $ \left\|\boldsymbol{H}-\bar{\boldsymbol{H}}\right\|$ can be given as
\begin{align}
    \left\|\bar{\boldsymbol{H}}-\boldsymbol{H}\right\|\le\tau\chi
\end{align}
where $\chi$ is a finite constant, and $\tau$ satisfies
\begin{align}
    \tau&=\left\|\boldsymbol{\Lambda}-\hat{\boldsymbol{\Lambda}}\right\|+\left\|\boldsymbol{M}\boldsymbol{R}^{-1}\boldsymbol{M}^{\text{T}}-\bar{\boldsymbol{M}}\boldsymbol{R}^{-1}\bar{\boldsymbol{M}}^{\text{T}}\right\| \notag\\
    &=\left\|\boldsymbol{\varepsilon_{\boldsymbol{\Lambda}}}\right\|+\left\|\boldsymbol{M}\boldsymbol{R}^{-1}\boldsymbol{M}^{\text{T}}-(\boldsymbol{M}+\boldsymbol{\varepsilon_{\varphi}}\boldsymbol{N})\boldsymbol{R}^{-1}(\boldsymbol{M}+\boldsymbol{\varepsilon_{\varphi}}\boldsymbol{N})^{\text{T}}\right\| \notag\\
    &\le\left\|\boldsymbol{\varepsilon_{\boldsymbol{\Lambda}}}\right\|+\left\|\boldsymbol{\varepsilon_{\varphi}}\right\|\Big(\left\|\boldsymbol{M}\boldsymbol{R}^{-1}\boldsymbol{N}^{\text{T}}\right\|+\left\|\boldsymbol{N}\boldsymbol{R}^{-1}\boldsymbol{M}^{\text{T}}\right\|\Big) \notag\\
    &\ \ \ +\left\|\boldsymbol{\varepsilon_{\varphi}}\right\|^{2}\left\|\boldsymbol{N}\boldsymbol{R}^{-1}\boldsymbol{N}^{\text{T}}\right\| \label{Theorem_2_5}
\end{align}

Therefore, the upper bound of $\left\|\boldsymbol{\mu}-\boldsymbol{\hat{\mu}}\right\|$ can be further derived as
\begin{align}
    &\left\|\boldsymbol{\mu}-\boldsymbol{\hat{\mu}}\right\|\notag\\
    &\le\left\|\boldsymbol{M}\boldsymbol{R}^{-1}\boldsymbol{M}^{\text{T}}\right\|\tau\chi\notag\\
    &\ \ \ +\left\|\boldsymbol{\varepsilon_{\varphi}}\right\|\Big(\left\|\boldsymbol{N}\boldsymbol{R}^{-1}\boldsymbol{M}^{\text{T}}\bar{\boldsymbol{H}}\right\|+\left\|\boldsymbol{M}\boldsymbol{R}^{-1}\boldsymbol{M}^{\text{T}}\bar{\boldsymbol{H}}\right\|\Big)\notag\\
    &\ \ \ +\left\|\boldsymbol{\varepsilon_{\varphi}}\right\|^{2}\left\|\boldsymbol{N}\boldsymbol{R}^{-1}\boldsymbol{N}^{\text{T}}\bar{\boldsymbol{H}}\right\|\notag\\
    &\le\left\|\boldsymbol{\varepsilon_{\boldsymbol{\Lambda}}}\right\|\left\|\boldsymbol{M}\boldsymbol{R}^{-1}\boldsymbol{M}^{\text{T}}\right\|\chi \notag\\
    &\ \ \ +\left\|\boldsymbol{\varepsilon_{\varphi}}\right\|\Big(\left\|\bar{\boldsymbol{H}}\right\|+\left\|\boldsymbol{M}\boldsymbol{R}^{-1}\boldsymbol{M}^{\text{T}}\right\|\chi\Big) \notag\\
    &\ \ \ \times\Big(\left\|\boldsymbol{N}\boldsymbol{R}^{-1}\boldsymbol{M}^{\text{T}}\right\|+\left\|\boldsymbol{M}\boldsymbol{R}^{-1}\boldsymbol{N}^{\text{T}}\right\|\Big)\notag\\
    &\ \ \ +\left\|\boldsymbol{\varepsilon_{\varphi}}\right\|^{2}\left\|\boldsymbol{N}\boldsymbol{R}^{-1}\boldsymbol{N}^{\text{T}}\right\|\Big(\left\|\bar{\boldsymbol{H}}\right\|+\left\|\boldsymbol{M}\boldsymbol{R}^{-1}\boldsymbol{M}^{\text{T}}\right\|\chi\Big).
 \label{eq_error_bound_in_mu_new}
\end{align}

From Eq.(\ref{eq_error_bound_in_mu_new}), it can be concluded that when $\left\|\boldsymbol{\varepsilon_{\boldsymbol{\Lambda}}}\right\|$ and $\left\|\boldsymbol{\varepsilon_{\varphi}}\right\|$ converge to 0, $\left\|\boldsymbol{\mu}-\boldsymbol{\hat{\mu}}\right\|$ converges to 0. In other words, the error of the closed-loop dynamics is limited by the representation errors of Koopman eigenpairs.

\begin{figure}[t] 
    \centering
    \includegraphics[width=8cm]{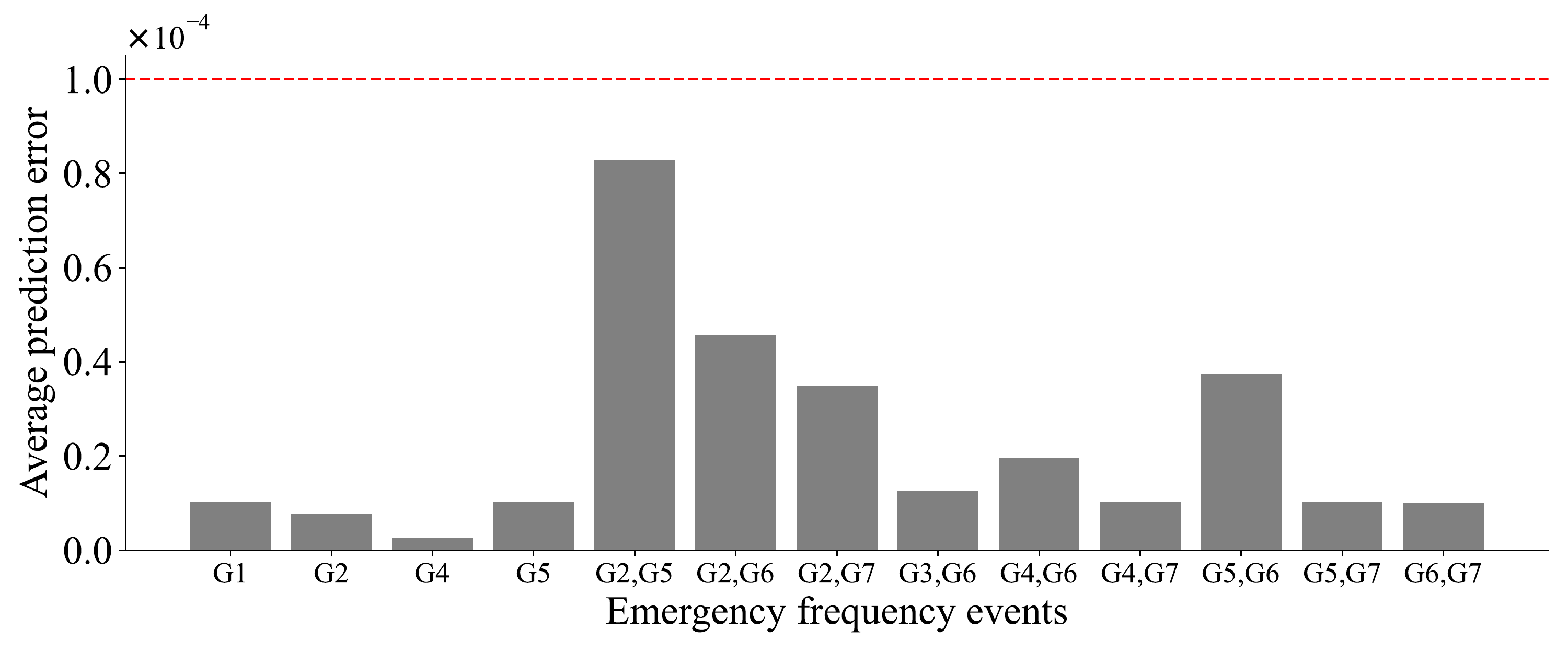}
    \caption{Average prediction errors of learnt Koopman eigenpairs on varying emergency frequency events}
    \label{Average_prediction_errors_of_learnt_Koopman_eigenpairs_on_varying_emergency_frequency_events}
\end{figure}

Here, we further discuss when does the equality holds in the inequality \eqref{eq_error_bound_in_mu_new}.

\vspace{0.2cm}
\textit{\textbf{Remark 1}}: In the deviation of the upper bound for $\left\|\boldsymbol{\mu}-\boldsymbol{\hat{\mu}}\right\|$, Eq.\eqref{eq_error_bound_in_mu}, Eq.\eqref{Theorem_2_5} and Eq.\eqref{eq_error_bound_in_mu_new} are based on the triangle inequality and the sub-multiplicative inequality of the Frobenius norm. For any two arrays $\boldsymbol{\mathcal{A}}$ and $\boldsymbol{\mathcal{B}}$, equality for the triangle inequality holds when the two arrays are linearly dependent, while equality for the sub-multiplicative inequality holds if and only if each row of $\boldsymbol{\mathcal{A}}$ and each column of $\boldsymbol{\mathcal{B}}$ are linearly dependent. 

\vspace{0.2cm}
\textit{\textbf{Remark 2}}: By Theorem 2.5 in Ref.\cite{bishop2018robustness}, a necessary condition for $\left\|\boldsymbol{H}-\bar{\boldsymbol{H}}\right\|$ to reach the upper bound in Eq.\eqref{Theorem_2_5} is the time of the Ricatti flow $t\rightarrow\infty$.

\vspace{0.2cm}
\textit{\textbf{Remark 3}}: $\left\|\boldsymbol{\mu}-\boldsymbol{\hat{\mu}}\right\|$ is often strictly lower than the upper bound derived in Eq.\eqref{eq_error_bound_in_mu_new}. One of the reasons is the equality conditions of the triangle inequality and the sub-multiplicative inequality in Eq.\eqref{eq_error_bound_in_mu}, Eq.\eqref{Theorem_2_5} and Eq.\eqref{eq_error_bound_in_mu_new} do not necessarily hold. For instance, for the equality in Eq.\eqref{eq_error_bound_in_mu} to hold, a necessary condition is each row of $\boldsymbol{M}\boldsymbol{R}^{-1}\boldsymbol{M}^{\text{T}}$ and each column of $\bar{\boldsymbol{H}}-\boldsymbol{H}$ are linearly dependent. However, the linear dependency is not guaranteed. Another reason is that it's unrealistic for an infinite control period in EFC. The gap between $\left\|\boldsymbol{\mu}-\boldsymbol{\hat{\mu}}\right\|$ and its upper bound will be further illustrated in the simulation results in Section \ref{Error_bound_analysis_with_eigenpairs_approximation_errors}.
\vspace{0.2cm}


In a conclusion, the robustness of EKEFC and upper bound of $\left\|\boldsymbol{\mu}-\boldsymbol{\hat{\mu}}\right\|$ promises the effectiveness of EKEFC when there exist representation errors in eigenpairs. In practical engineering, it's necessary for EKEFC to adapt to complicated online operational contexts in EFC such as unknown time-delay measurements and the deadzone setting. The robustness of EKEFC and the upper bound of $\left\|\boldsymbol{\mu}-\boldsymbol{\hat{\mu}}\right\|$ help to address the problems of these uncertainties or perturbation in the system dynamics.

\begin{figure}[t] 
    \centering
    \includegraphics[width=6.5cm]{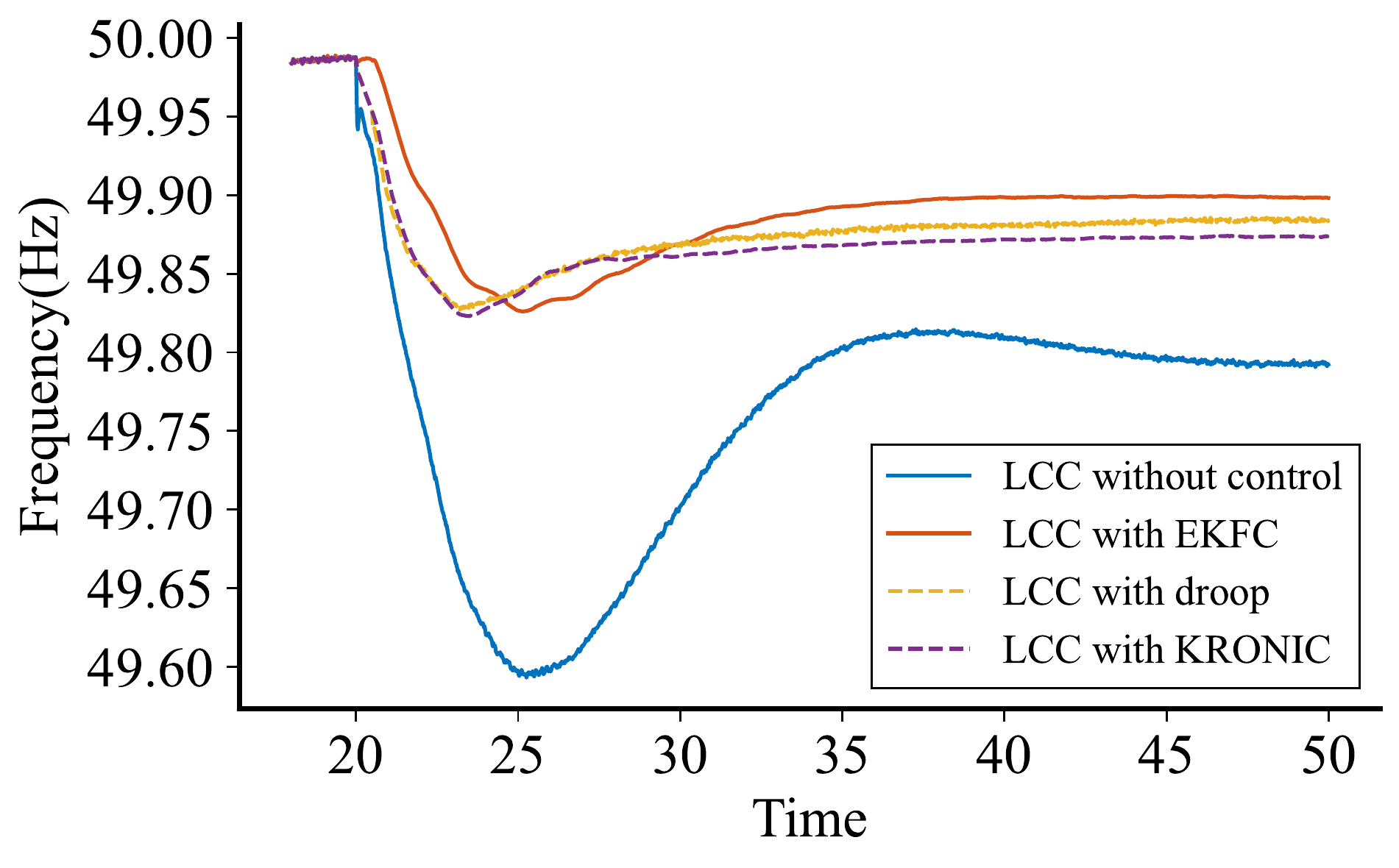}
    \caption{Frequencies of the AC main system when G6 trip at 20s}
    \label{fig_Frequencies_of_the_AC_main_system_when_G6_trip_at_20s}
\end{figure}

\section{Case Study} \label{case_study}
In this section, the effectiveness of EKEFC is illustrated by a case study on the CloudPSS platform \cite{SONG20201611}, \cite{CloudPSS}. All of the following tests are conducted on PCs with Intel Xeon W-2255 processor, 3.70 GHz primary frequency, and 128GB memory.
\vspace{-0.2cm}
\subsection{Test System and Datasets} \label{Test_System_Description}
The MIDC test system is a modified IEEE New England system combining the CIGRE HVDC benchmark systems \cite{faruque2005detailed} used in Ref.\cite{2021Optimal}. The full electromagnetic transient (EMT) model of the test system is built on the CloudPSS platform \cite{8582334}. The main AC system is connected with four ±660 kV monopolar 12-pulse LCC-HVDC systems. The system capacity is 4000MW.

To identify eigenpairs for frequency dynamics in MIDC systems, we set generator G6 trip at the time of 20s, which causes a 530 MW power imbalance. With the control input $\textbf{\textit{u}}=\boldsymbol{0}$, data of $\boldsymbol{x}=\{\delta_h,\omega_i,{p}_i^{dc}\ |\ h\in \mathcal{N}_\mathcal{G}\cup\mathcal{N}_\mathcal{D}, i\in\mathcal{N}_\mathcal{D}\}$ were collected in the time span of $20\sim50$s with a rate of 100Hz (3000 time points in total).

To identify the control matrix $\boldsymbol{B}$, data of $\boldsymbol{x}$ from the MIDC system with random control input should also be collected. Uniform-distributed numbers were generated as the control input in the time span of $20\sim50$s with a rate of 100Hz. Since DC power reference regulation amount is constrained to be $-20\%$ and $+10\%$ of the nominal transmission power of each LCC-HVDC, the uniform distribution is limited on $[-0.2\ p.u., +0.1\ p.u.]$. Data of $\textbf{\textit{x}}^u$ were collected in the time span of $20\sim50$s with a rate of 100Hz. 
Based on the above settings, we obtain the following results.
\vspace{-0.2cm}
\subsection{Obtained Linear Representations} \label{Obtained_Linear_Representation}
\subsubsection{Library Construction}
For library setting, a polynomial basis up to the second order, trigonometric terms of $\boldsymbol{x}$, trigonometric transform of subtraction between any two rotor angles are employed. 

Considering the dynamics of a MIDC system given as Eq.\eqref{eq_theta}-Eq.\eqref{eq_droop_control_equation}, there are totally $n=22$ state variables in the test system, of which 7 are the rotor angles of $\mathcal{N}_\mathcal{G}$, 11 are the frequencies deviation from the nominal frequency at $\mathcal{N}_\mathcal{G}$ and $\mathcal{N}_\mathcal{D}$, and 4 are the transmission power of LCC-HVDC. Based on library construction method in Section \ref{Integrating_nonlinearities_of_the_MIDC_system}, there are 318 basis functions in total, of which 275 are polynomial terms, 22 are trigonometric transform of $\textbf{\textit{x}}$, and 21 are trigonometric transform of subtraction between any two rotor angles.
\subsubsection{Library Subsampling and Ensemble Learning for Eigenfunctions}
According to the library subsampling method in Section \ref{Library_Subsampling_and_Ensemble_Learning_for_Eigenpairs}, $D=2^{5-1}=16$ subsets are constructed in total.

Without the library bagging technique, the only verified eigenfunction with a prediction error under $1e^{-4}$ is $\cos\omega_{36}$ with an eigenvalue of $-1.19e^{-4}$. Verified eigenpairs obtained by introducing the library bagging technique are listed in TABLE.\ref{Eigenpairs_learnt_in_EKEFC} in Appendix \ref{section_B}. After the library bagging technique is introduced, the number of verified eigenfunctions with a prediction error under $1e^{-4}$ increases to 16. 

Considering realistic communication infrastructure, we design distributed control with partial measurements, in which only system frequency $\omega_i$ and $p_i^{dc}\ (i\in \mathcal{N}_\mathcal{D})$ are known. Therefore, the eigenpairs \#1-9 can be further selected to form $\boldsymbol{\varphi}_i$ in Eq.\eqref{Koopman_based_LQR_for_HVDC} for LCC-HVDC $i$, since the others are expressed explicitly in terms of rotor angles.

\subsubsection{Generality of Eigenpairs}
\label{Generalization_Capability_of_Eigenfunctions}
Uncertain emergency faults are unavoidable conditions in practical online operational contexts. To identify eigenpairs for frequency dynamics in MIDC systems, trajectories of $\textbf{\textit{x}}$ after the trip of G6 are collected. To illustrate the generality of the learnt Koopman eigenpairs, we further calculate prediction errors of the verified Koopman eigenpairs on trajectories of other trip events. If the prediction errors are still small (under the threshold $1e^{-4}$), then the generalization capability of the learnt Koopman eigenpairs can be verified. Other emergency faults considered include:

i) trip of another generator;

ii) trip of two generators at the same time.

We traverse all emergency frequency events in the above two cases in the MIDC test system. There are 7 generators in the system, thus $(7-1)+\text{C}_7^2=27$ scenarios can be obtained. Data of $\boldsymbol{x}$ were collected in the time span of $20\sim50$s with a rate of 100Hz in each scenario. Since EKEFC is designed to improve the system frequency stability, we exclude the scenarios where voltage instability or angle instability occurs. We calculate prediction errors of the eigenpairs \#1-9 on $\boldsymbol{x}$ collected in each scenario. The average prediction error of the eigenpairs in each scenario are given in Fig.\ref{Average_prediction_errors_of_learnt_Koopman_eigenpairs_on_varying_emergency_frequency_events}. As shown in Fig.\ref{Average_prediction_errors_of_learnt_Koopman_eigenpairs_on_varying_emergency_frequency_events}, prediction errors are still small even when different trip events occur. Therefore, the generality of the eigenpairs is verified, although the eigenpairs are obtained on limited datasets.

\begin{figure}[t] 
    \centering
    \includegraphics[width=9cm]{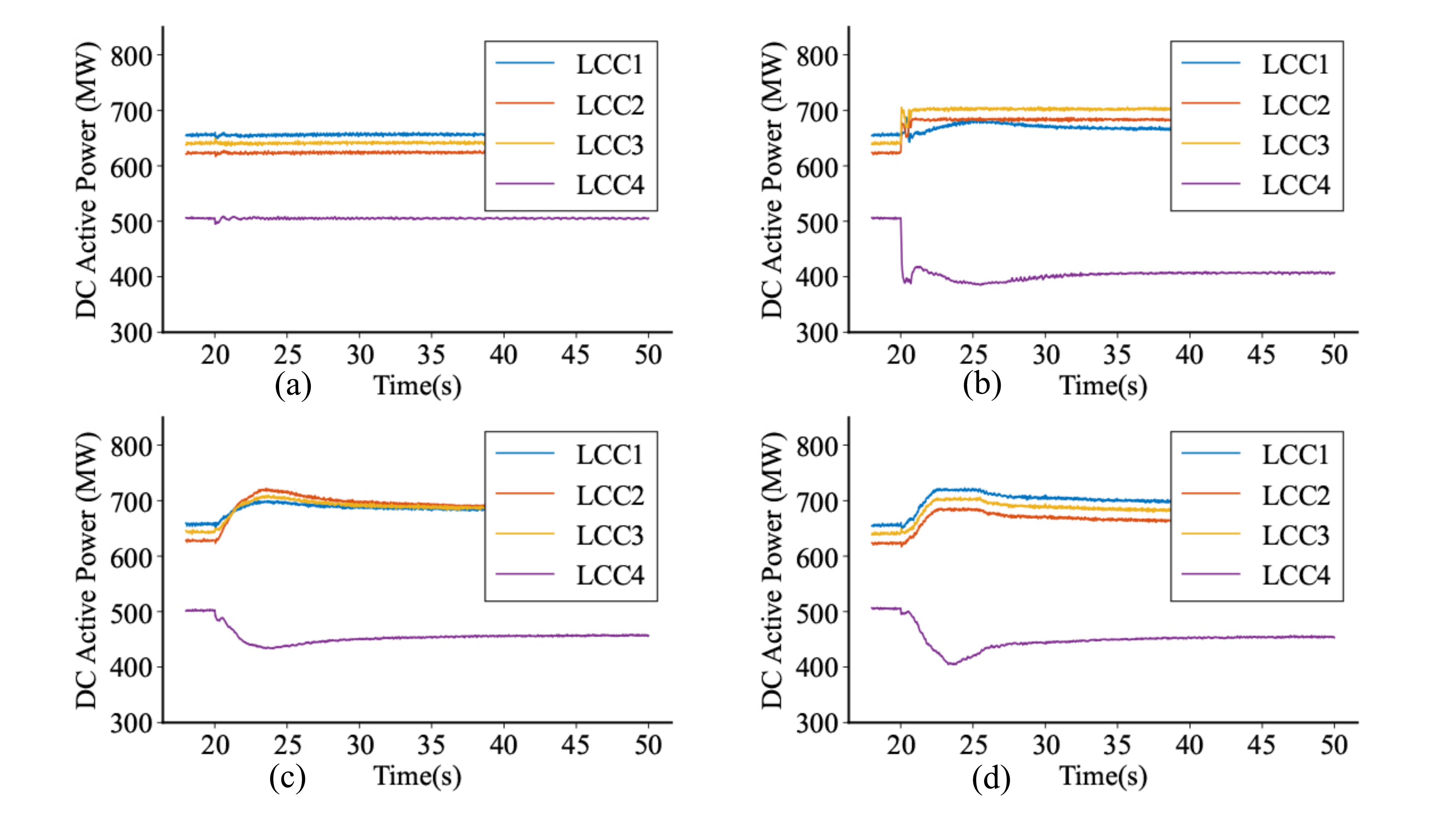}
    \caption{Active  powers  of  the LCC-HVDCs. (a) Subcase (1). (b) Subcase (2). (c) Subcase (3). (d) Subcase (4).}\label{fig_Active_powers_of_the_LCC-HVDCs}
\end{figure}
\vspace{-0.2cm}
\subsection{Effectiveness of EKEFC} \label{Effectiveness_of_Coordinated_Droop_Based_EFC}

To demonstrate the effectiveness of EKEFC, the active power reference of LCC-HVDC is determined adaptively to participate in frequency control by EKEFC after generator G6 tripped at the time of 20s. $\boldsymbol{Q}_i$ is a diagonal weight matrix, which has 1 as the entry when the corresponding eigenfunction is expressed only in terms of frequencies, and 0 for other eigenfunctions. The weight matrice $R=2e^{-6}$. In our simulation, Eq.\eqref{Koopman_based_LQR_for_HVDC} can be solved within 0.02s. Therefore, the control period $\Delta t_1$ is set as 0.1s. 

To assess the performance of EKEFC, it's compared with P-f droop based EFC, a typical frequency control strategy, and KRONIC, which is a representative Koopman based control design method. Therefore, the following four subcases are compared: (1) The LCC-HVDCs have no control designed to provide frequency support for the system. (2) All LCC-HVDCs have EKEFC as the EFC strategy. (3) All LCC-HVDCs have droop control with optimal coefficients calculated in Ref.\cite{2021Optimal}. (4) All LCC-HVDCs have KRONIC as the EFC strategy. The results of frequencies of the AC main system are displayed in Fig.\ref{fig_Frequencies_of_the_AC_main_system_when_G6_trip_at_20s}. The active powers of LCC-HVDCs are shown in Fig.\ref{fig_Active_powers_of_the_LCC-HVDCs}.

As shown in Fig.\ref{fig_Frequencies_of_the_AC_main_system_when_G6_trip_at_20s}, the AC main system frequency reduces to  49.6 Hz at approximately 25 s. However, for systems with capacity above 3000MW, the allowable frequency deviation is $50\pm0.2$Hz. By adopting EKEFC strategy, the system frequency stabilizes at approximately 49.88 Hz at approximately 35 s. Compared with subcase (1), subcase (2) has a shorter transient time, and the steady-state frequency is closer to the nominal frequency. Moreover, the proposed EKEFC method regulates the frequency nadir, which increase from 49.6Hz to 49.83Hz. The frequencies during the transient time are within the allowable frequency deviation. Thus, the proposed EFFC strategy is effective.

In Fig.\ref{fig_Frequencies_of_the_AC_main_system_when_G6_trip_at_20s}, the system frequency of subcase (2) is higher than that in subcase (3). This is because eigenfunctions predict system dynamics globally and solve the open-loop optimization problem over an infinite time horizon. However, the droop control law is not able to predict frequency over time and is restricted to be proportional to frequency deviation. Moreover, the system frequency of subcase (2) is higher than that in subcase (4). This is because more verified eigenfunctions can be obtained to describe system dynamics in EKEFC, so that more control objects can be included in the cost function in Eq.\eqref{cost_function}.

As shown in Fig.\ref{fig_Active_powers_of_the_LCC-HVDCs}, in subcase (1), the emergency frequency regulation can only rely on the generators’ primary droop, but the power adjustment speed of generators is relatively slow. In subcases (2), (3) and (4), the fast power adjustability of the LCC-HVDC systems is utilized to provide considerable power support and relieve the frequency modulation pressure of the generators. By comparing Fig.\ref{fig_Active_powers_of_the_LCC-HVDCs}(b) and Fig.\ref{fig_Active_powers_of_the_LCC-HVDCs}(c), we also see that during $20\sim23$s, subcase (2) provides large power support at the moment emergency faults occur, while in subcase (3) the DC power gradually increases as the frequency decreases.

\begin{figure}[t] 
    \centering
    \includegraphics[width=9cm]{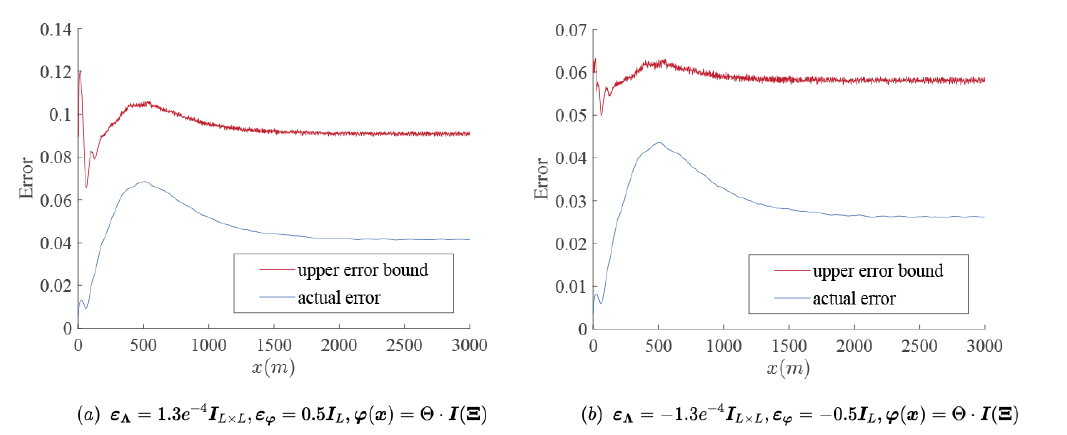}
    \caption{Estimated upper error bound and actual error}\label{fig_Estimated_upper_error_bound_and_actual_error}
\end{figure}
\begin{figure}[t] 
    \centering
    \includegraphics[width=8.5cm]{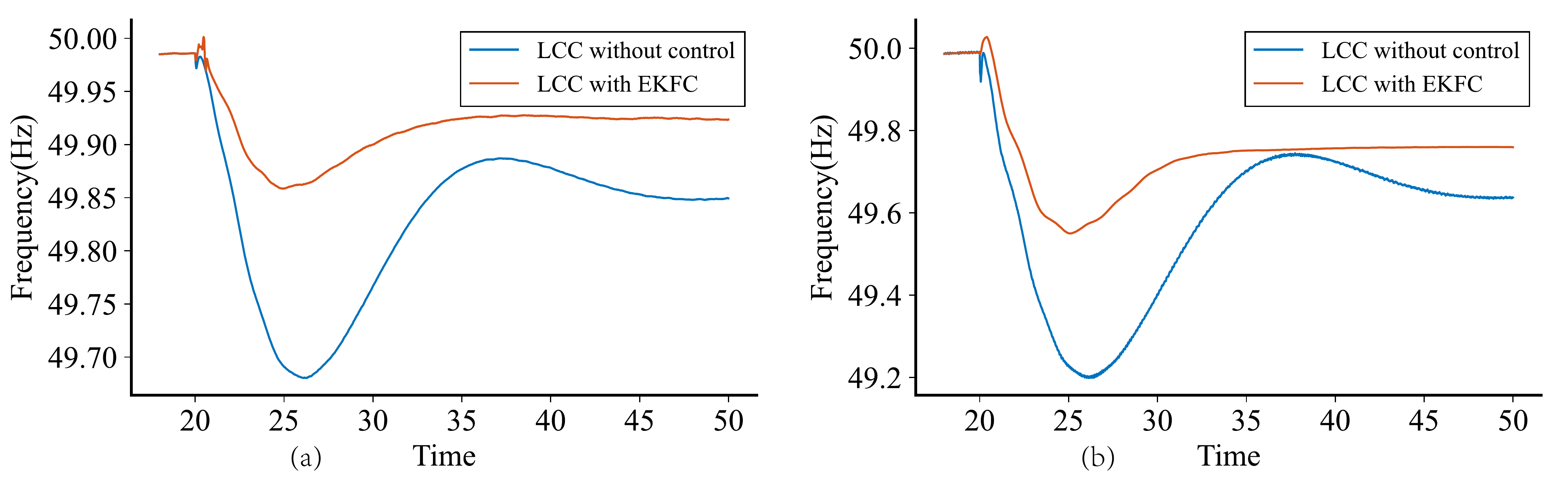}
    \caption{Frequencies of the AC main system (a)when G4 trip at 20s. (b)when  G4,  G6  trip  at  20s}\label{Fig_5_new}
\end{figure}
\vspace{-0.2cm}
\subsection{Error bound analysis with eigenpairs approximation errors}
\label{Error_bound_analysis_with_eigenpairs_approximation_errors}
In Section \ref{Robustness_Analysis_and_Error_Estimation}, an expression for the upper error bound of $\boldsymbol{\mu}$ is provided. To demonstrate the actual error of $\boldsymbol{\mu}$ is strictly below the upper error bound, we consider an analytical example in Ref.\cite{Kaiser_2021} and the MIDC example in Section \ref{Test_System_Description}.

In the analytical example, a closed and finite-dimensional Koopman approximation exists. $\boldsymbol{\varepsilon_{\Lambda}}$ and $\boldsymbol{\varepsilon_{\varphi}\psi(x)}$ can be set to model possible errors in the representation of the Koopman operator. The results show that the actual errors along a given trajectory are strictly lower than the upper error bounds, validating our estimation for the error bound in Eq.\eqref{eq_error_bound_in_mu_new}. For more details, see Appendix \ref{section_A}.

In MIDC examples, since the accurate Koopman operator is inaccessible, we assume that the verified eigenpairs obtained in Section \ref{Obtained_Linear_Representation} are the accurate ones. To demonstrate how the representation errors of eigenpairs influence the control effect of EKEFC, $\boldsymbol{\varepsilon}_{\boldsymbol{\Lambda}}$ and $\boldsymbol{\varepsilon}_{\boldsymbol{\varphi}}\boldsymbol{\psi}(\textbf{\textit{x}})$ defined in Section \ref{Robustness_Analysis_and_Error_Estimation} can be artificially given. Then the system dynamics with EKEFC of accurate eigenpairs and of eigenpairs with representation errors can be simulated. Subsequently, both sides of the inequality Eq.(\ref{eq_error_bound_in_mu_new}) can be calculated. If $\left\|\boldsymbol{\mu}-\boldsymbol{\hat{\mu}}\right\|$ is strictly lower than the estimated upper error bound, then the inequality Eq.(\ref{eq_error_bound_in_mu_new}) can be verified. 

To simulate representation errors of eigenpairs, we set $\boldsymbol{\varepsilon_{\varphi}}=\epsilon \Bbb I_{L}$ and $ \psi(\textbf{\textit{x}})=\boldsymbol{I(\Xi)}\cdot\Theta$, where $\epsilon$ is a real number, $\Bbb I_{L}$ represents an $L\times L$ identity matrix, $\boldsymbol{I(\Xi)}$ is a $P\times L$ matrix in which $I_{ij}(\boldsymbol{\Xi})=1$ if $\Xi_{ij}\neq0$, else $I_{ij}(\boldsymbol{\Xi})=0$.

Fig.\ref{fig_Estimated_upper_error_bound_and_actual_error} demonstrates two examples with different error settings. The actual errors and the upper error bounds are calculated on the trajectory of the closed-loop system in subcase (2) with accurate eigenpairs.

The results show that the actual errors along the given trajectory are strictly lower than the estimated upper error bounds, validating our estimation for the error bound in Eq.\eqref{eq_error_bound_in_mu_new}. As discussed in Section \ref{Robustness_Analysis_and_Error_Estimation}, $\left\|\boldsymbol{\mu}-\boldsymbol{\hat{\mu}}\right\|$ is often strictly lower than the upper bound derived in Eq.\eqref{eq_error_bound_in_mu_new}. Thus, we can conclude that when $\left\|\boldsymbol{\varepsilon_{\Lambda}}\right\|$ and $\left\|\boldsymbol{\varepsilon_{\varphi}}\right\|$ are finite, $\left\|\boldsymbol{\mu}-\boldsymbol{\hat{\mu}}\right\|$ is also finite.

\begin{figure*}[t] 
    \centering
    \includegraphics[width=17cm]{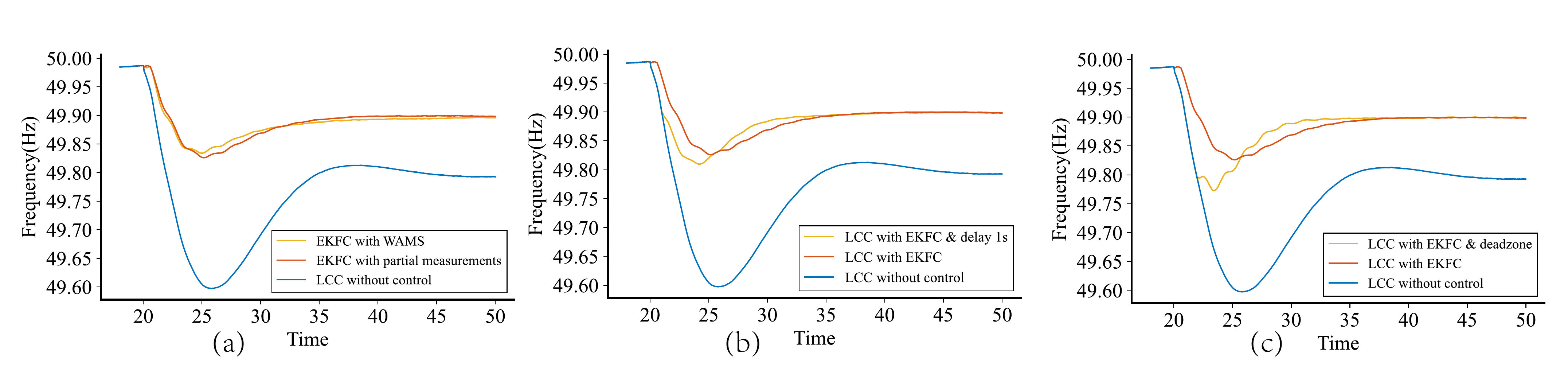}
    \caption{Frequencies of the AC main system (a) when $\omega_s\ (s\in\mathcal{N}_\mathcal{D})$ is measurable at LCC-HVDC $i$. (b) when measurements has 1s time delay. (c) when EKEFC has 49.80Hz deadzone.}\label{Fig_6_new_1125}
\end{figure*}
\vspace{-0.2cm}
\subsection{Adaptability test}
Uncertain emergency faults, time-delay measurements are unavoidable in practical operation. Moreover, in engineering practice, a dead zone setting for EKEFC is necessary. These realistic conditions may weaken the control effect. Therefore, the adaptability of EFKC to such realistic conditions is examined here. Note that adaptability of EFKC to uncertain emergency faults benefits from generality of the obtained eigenpairs on different faults, while adaptability of EFKC to time-delay measurements and the dead zone setting benefits from the robustness to eigenpairs with representation errors.

\subsubsection{Unknown emergency events}
In Section V-A, trajectories of $\textbf{\textit{x}}$ after the trip of G6 are collected to construct system dynamics in eigenfunction coordinates. Here, varying trip events are considered to examine the adaptability of EKEFC. The trip events considered are the same as in Section \ref{Generalization_Capability_of_Eigenfunctions}.

Fig.\ref{Fig_5_new} illustrates the adaptability of EKEFC to unknown emergency frequency events. Note that EKEFC still achieves satisfactory control performances in such totally unknown events. Results of more adaptbility tests of EKEFC to unknown emergency frequency events are given in TABLE.\ref{Adaptbility_tests_of_EKEFC_to_unknown_emergency_frequency_events} in Appendix \ref{section_B}.

\subsubsection{Partial measurements}
In Section \ref{Koopman_Operator_Based_Emergency_Frequency_Control_Strategy}, we assume that the frequency dynamics in any node or generator are the same. Therefore, we let $\omega_s\ (s\in\mathcal{N}_\mathcal{D})$ in ${\boldsymbol{\varphi}}_i$ take the value of $\omega_i$ in the optimal control strategy calculation for LCC-HVDC $i$. The influence of neglecting spatial-temporal characteristics of frequencies is illustrated in Fig.\ref{Fig_6_new_1125}(a). The results show that partial measurements causes the frequency nadir to be about 0.01Hz lower but has no effect on the steady-state frequency.

\subsubsection{Time-delay of measurements}
Time-delay is unavoidable due to communication latency or control strategy computation. In practical grid, the time delay is often below 150ms\cite{9024098}. Here, to test the performance of EKEFC, we assume that the measurements have an 1s delay. The results are illustrated in Fig.\ref{Fig_6_new_1125}(b).

The results show that time-delay of measurements causes the frequency nadir to be about 0.02Hz lower but has no effect on the steady-state frequency.
\subsubsection{A deadzone setting}
In engineering practice, a dead zone setting for EKEFC is necessary. When the system frequency changes due to some faults, the frequency limitation of the dead zone is utilized to determine whether there is an emergency and whether to enable EKEFC. In this paper, we assume that a frequency deviation limitation is used to set the deadzone. The influence of the deadzone setting of EKEFC is illustrated in Fig.\ref{Fig_6_new_1125}(c).

The results show that after EKEFC is triggered by the dead zone setting, the frequency of the AC main system is stabilized soon. During the transient frequency process, the dead zone setting causes the frequency nadir to be about 0.2Hz lower than the frequency nadir in subcase (2). However, the dead zone setting has no effect on the steady-state frequency.

The adaptability of EKEFC benefits from generality of obtained Koopman eigenpairs and the robustness to eigenpairs approximation errors. As long as Eq.(\ref{eq_tilde_P}) is satisfied, EKEFC can adapt to conditions not included in the initial dataset. 

The above results imply that EKEFC does not necessarily rely on access to massive datasets of different scenarios and is potential to enable the control of nonlinear systems even when limited scenarios are considered.

\section{Conclusion}
In this paper, the discovery of linear representations of nonlinear MIDC system dynamics is developed based on Koopman operator theory. Applying the linear representations, a fully data-driven dynamic optimal control method EKEFC is proposed for LCC-HVDCs to participate in the system frequency regulation service. Furthermore, an error bound of the closed-loop dynamics with consideration of Koopman operator approximation errors is estimated. The case study demonstrates the effectiveness of EKEFC on providing frequency support. Moreover, actual errors are proved to be strictly lower than the upper bound estimated for the closed-loop dynamics. Simulation results show that EKEFC adapts to practical conditions such as uncertain emergency faults and time-delay measurements. Furthermore, a dead zone setting for EKEFC has no effect on the steady-state frequency. The above results indicate that EKEFC does not necessarily rely on access to massive datasets of different scenarios and is potential to enable the control of nonlinear systems even when limited scenarios are considered.
 
\bibliographystyle{IEEEtran}
\bibliography{ref}

\appendices

\section{} \label{section_A}
For most practical dynamic systems, accurate Koopman eigenpairs are often inaccessible since representation errors of Koopman eigenpairs are often unavoidable. Here, to demonstrate the actual error is strictly bounded by the error bound estimated in Eq.\eqref{eq_error_bound_in_mu_new}, we consider an analytical example given as Eq.\eqref{an_analytical_example}, where a closed and finite-dimensional Koopman approximation exists.

\begin{equation}
  \frac{d}{dt}\left[
  \begin{gathered}
  x_1\\
  x_2\\
  \end{gathered}
  \right]=\left[
  \begin{gathered}
  \mu_1x_1\\
  \mu_2(x_2-x_1^2)\\
  \end{gathered}
  \right]+\boldsymbol{B}u. \label{an_analytical_example}
\end{equation}

This nonlinear system can be embedded in a higher-dimensional space $(y_1,y_2,y_3) = (x_1,x_2,x^2_1)$ where the unforced dynamics form a closed linear system in a Koopman-invariant subspace:
\begin{equation}
  \frac{d}{dt}\left[
  \begin{gathered}
  y_1\\
  y_2\\
  y_3\\
  \end{gathered}
  \right]=\begin{aligned}
    &\left [
    \begin{matrix}
        \mu_1&0&0\\
        0&\mu_2&-\mu_2\\
        0&0&2\mu_1\\
    \end{matrix}
        \right ]\left[
  \begin{gathered}
  y_1\\
  y_2\\
  y_3\\
  \end{gathered}
  \right]+\left [
    \begin{matrix}
        1&0\\
        0&1\\
        2y_1&0\\
    \end{matrix}
        \right ]\boldsymbol{B}u.
\end{aligned}
\end{equation}

Koopman eigenfunctions of the unforced system are $\varphi_{\mu_1}=x_1, \varphi_{\mu_2}=x_2-bx_1^2$ with $b=\frac{\mu_2}{\mu_2-2\mu_1}$ corresponding to the eigenvalue $\mu_1$ and $\mu_2$, respectively.

\begin{figure}[h] 
    \centering
    \includegraphics[width=6cm]{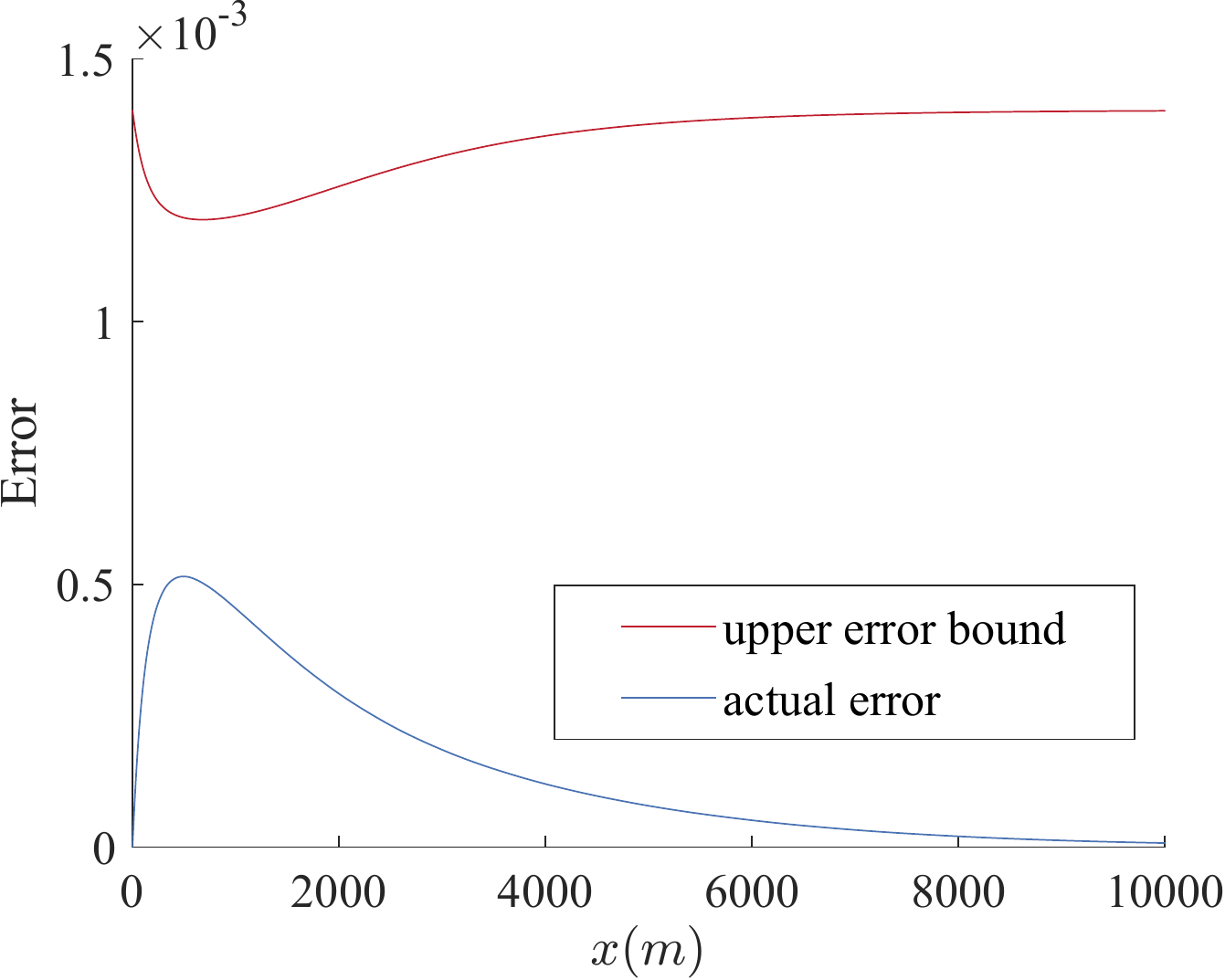}
    \caption{Estimated upper error bounds and actual errors of an controlled system in which a closed Koopman approimation exists}\label{Estimated_upper_error_bounds_and_actual_errors_of_an_controlled_system_in_which_a_closed_Koopman_approimation_exists}
\end{figure}

Here, we set $\mu_1=-0.1$ and $\mu_2=-1$. Errors in the representation of $\mu_1$ and $\mu_2$ are simulated as $\epsilon_{\mu_1}=-1e^{-3}$ and $\epsilon_{\mu_2}=-1e^{-3}$. The errors in the representation of eigenfunctions are simulated as $\varepsilon_{\varphi}\psi(\textbf{\textit{x}})=-0.2816\Bbb I_{3}\cdot(0,x_1^{2},0)$, where $\Bbb I_{3}$ is a $3\times 3$ identity matrix. The results are illustrated in Fig.\ref{Estimated_upper_error_bounds_and_actual_errors_of_an_controlled_system_in_which_a_closed_Koopman_approimation_exists}.


\section{} \label{section_B}
\begin{table*}[t]
\renewcommand\arraystretch{1.5}
\footnotesize
\caption{Eigenpairs learnt in EKEFC}
\label{Eigenpairs_learnt_in_EKEFC}
\begin{tabular}{llllllllllllllllllllllllllll}
\hline
\multicolumn{1}{c}{\multirow{2}{*}{No.}} &
  \multicolumn{6}{c}{\multirow{2}{*}{Eigenvalue}} &
  \multicolumn{20}{c}{\multirow{2}{*}{Eigenfunction}} &
  \multicolumn{1}{c}{\multirow{2}{*}{\begin{tabular}[c]{@{}c@{}}Prediction \\ error\end{tabular}}} \\
\multicolumn{1}{c}{} &
  \multicolumn{6}{c}{} &
  \multicolumn{20}{c}{} &
  \multicolumn{1}{c}{} \\ \hline
1 &
  \multicolumn{6}{l}{$-1.26e^{-9}$} &
  \multicolumn{20}{l}{$0.175P^{dc}_{33}+0.141\sin P^{dc}_{33}+1.000\cos P^{dc}_{33}$} &
  $1.95e^{-7}$ \\
2 &
  \multicolumn{6}{l}{$-6.23e^{-9}$} &
  \multicolumn{20}{l}{$0.190P^{dc}_{35}+0.135\sin P^{dc}_{35}+1.000\cos P^{dc}_{35}$} &
  $2.19e^{-7}$ \\
3 &
  \multicolumn{6}{l}{$-1.58e^{-9}$} &
  \multicolumn{20}{l}{$0.139P^{dc}_{8}+0.117\sin P^{dc}_{8}+1.000\cos P^{dc}_{8}$} &
  $3.43e^{-7}$ \\
4 &
  \multicolumn{6}{l}{$2.91e^{-8}$} &
  \multicolumn{20}{l}{$0.198P^{dc}_{31}+0.135\sin P^{dc}_{31}+1.000\cos P^{dc}_{31}$} &
  $7.33e^{-7}$ \\
5 &
  \multicolumn{6}{l}{$3.28e^{-8}$} &
  \multicolumn{20}{l}{$1.000\cos\omega_{34}$} &
  $5.07e^{-6}$ \\
6 &
  \multicolumn{6}{l}{$-2.2e^{-7}$} &
  \multicolumn{20}{l}{$1.000\cos\omega_{30}$} &
  $8.42e^{-6}$ \\
7 &
  \multicolumn{6}{l}{$-1.39e^{-11}-5.8e^{-10}j$} &
  \multicolumn{20}{l}{$0.017P^{dc}_{8}+1.000\cos\omega_{38}$} &
  $1.09e^{-5}$ \\
8 &
  \multicolumn{6}{l}{$-7.35e^{-7}$} &
  \multicolumn{20}{l}{$1.000\cos\omega_{32}$} &
  $1.15e^{-5}$ \\
9 &
  \multicolumn{6}{l}{$-2.88e^{-6}$} &
  \multicolumn{20}{l}{$1.000\cos\omega_{36}$} &
  $4.25e^{-5}$ \\
11 &
  \multicolumn{6}{l}{$-7.84e^{-8}$} &
  \multicolumn{20}{l}{$0.698\delta_{34}+1.000\cos\delta_{34}$} &
  $5.27e^{-5}$ \\
12 &
  \multicolumn{6}{l}{$1.94e^{-10}-1.13e^{-10}j$} &
  \multicolumn{20}{l}{$1.000\sin\delta_{32}+(0.832+0.001j)\cos\delta_{32}$} &
  $6.94e^{-5}$ \\
\multirow{2}{*}{13} &
  \multicolumn{6}{l}{\multirow{2}{*}{$-1.22e^{-10}-9.14e^{-11}j$}} &
  \multicolumn{20}{l}{\multirow{2}{*}{\begin{tabular}[c]{@{}l@{}}$(0.002-0.001j)\delta_{32}+(0.008-0.008j)\delta_{34}+1.000\sin\delta_{32}$\\ $+(0.840+0.001j)\cos\delta_{32}+(0.014-0.019j)\sin(\delta_{32}-\delta_{34})+(0.001-0.001j)\sin(\delta_{34}-\delta_{36})$\end{tabular}}} &
  \multirow{2}{*}{$1.24e^{-4}$} \\
 &
  \multicolumn{6}{l}{} &
  \multicolumn{20}{l}{} &
   \\
14 &
  \multicolumn{6}{l}{$1.40e^{-9}$} &
  \multicolumn{20}{l}{$0.002\delta_{34}+1.000cos\delta_{30}+0.005sin(\delta_{34}-\delta_{36})$} &
  $4.76e^{-4}$ \\
15 &
  \multicolumn{6}{l}{$4.5e^{-9}$} &
  \multicolumn{20}{l}{$-0.033\delta_{34}-0.001\sin(\delta_{34}-\delta_{36})+1.000\cos(\delta_{32}-\delta_{34})$} &
  $7.06e^{-4}$ \\
16 &
  \multicolumn{6}{l}{$-6.68e^{-9}$} &
  \multicolumn{20}{l}{$-0.002\delta_{36}+0.065\sin(\delta_{34}-\delta_{36})+1.000\sin(\delta_{38}-\delta_{30})$} &
  $7.57e^{-4}$ \\ \hline
\end{tabular}
\end{table*}

\begin{table}[h]
\centering
\caption{Adaptbility tests of EKEFC to unknown emergency frequency events}
\label{Adaptbility_tests_of_EKEFC_to_unknown_emergency_frequency_events}
\footnotesize
\renewcommand{\arraystretch}{1.2}
\setlength{\tabcolsep}{3pt}
\begin{tabular}{lcccc}
\bottomrule
\multicolumn{1}{c}{}                   & \multicolumn{2}{c}{G1}   & \multicolumn{2}{c}{G2}                                     \\ \cline{2-5} 
\multicolumn{1}{c}{\multirow{-2}{*}{}} & nadir    & steady-state  & nadir                       & steady-state                 \\ \hline
LCC without control                    & 49.78    & 49.82         & 49.63                      & 49.82                       \\
LCC with EKEFC                          & 49.91   & 49.99       & 49.83                       & 49.91                       \\ \hline
\multicolumn{1}{c}{}                   & \multicolumn{2}{c}{G4}   & \multicolumn{2}{c}{G5}                                     \\ \cline{2-5} 
\multicolumn{1}{c}{\multirow{-2}{*}{}} & nadir    & steady-state  & nadir                       & steady-state                 \\ \hline
LCC without control                    & 49.74   & 49.88        & 49.50                        & 49.74                       \\
LCC with EKEFC                          & 49.93   & 49.96       & 49.75                       & 49.86                      \\ \hline
\multicolumn{1}{c}{}                   & \multicolumn{2}{c}{G2, G5} & \multicolumn{2}{c}{G2, G6}                                   \\ \cline{2-5} 
\multicolumn{1}{c}{\multirow{-2}{*}{}} & nadir    & steady-state  & nadir                       & steady-state                 \\ \hline
LCC without control                    & 49.16   & 49.52        & 49.19                       & 49.56                       \\
LCC with EKEFC                          & 49.40    & 49.70         & 49.50                        & 49.73                       \\ \hline
\multicolumn{1}{c}{}                   & \multicolumn{2}{c}{G2, G7} & \multicolumn{2}{c}{G3, G6}                                   \\ \cline{2-5} 
\multicolumn{1}{c}{\multirow{-2}{*}{}} & nadir    & steady-state  & nadir                       & steady-state                 \\ \hline
LCC without control                    & 49.50     & 49.73         & {$\bullet$ 49.50} & {$\bullet$ 49.73} \\
LCC with EKEFC                          & 49.72   & 49.84       & {$\bullet$ 49.70} & {$\bullet$ 49.82} \\ \hline
\multicolumn{1}{c}{}                   & \multicolumn{2}{c}{G4, G6} & \multicolumn{2}{c}{G4, G7}                                   \\ \cline{2-5} 
\multicolumn{1}{c}{\multirow{-2}{*}{}} & nadir    & steady-state  & nadir                       & steady-state                 \\ \hline
LCC without control & {$\bullet$ 49.20}  & {$\bullet$ 49.66} & {$\#$ 49.53}  & {$\#$ 49.80} \\
LCC with EKEFC       & {$\bullet$ 49.55} & {$\bullet$ 49.76}  & {$\#$ 49.66} & {$\#$ 49.84}  \\ \hline
\multicolumn{1}{c}{}                   & \multicolumn{2}{c}{G5, G6} & \multicolumn{2}{c}{G5, G7}                                   \\ \cline{2-5} 
\multicolumn{1}{c}{\multirow{-2}{*}{}} & nadir    & steady-state  & nadir                       & steady-state                 \\ \hline
LCC without control                    & 49.07    & 49.52       & 49.93                     & 49.65                        \\
LCC with EKEFC                          & 49.41    & 49.7          & 49.96                      & 49.79                        \\ \hline
\multicolumn{1}{c}{}                   & \multicolumn{2}{c}{G6, G7} &                             &                              \\ \cline{2-3}
\multicolumn{1}{c}{\multirow{-2}{*}{}} & nadir    & steady-state  &                             &                              \\ \hline
LCC without control                    & 49.50     & 49.73       &                             &                              \\
LCC with EKEFC                          & 49.76   & 49.85        &                             &                              \\ 
\bottomrule
\multicolumn{4}{l}{\small $\bullet$ denotes the weight matrices are $\boldsymbol{Q}=1, \boldsymbol{R}=2e^{-5}$.}\\
\multicolumn{4}{l}{\small $\#$ denotes the weight matrices are $\boldsymbol{Q}=1, \boldsymbol{R}=2e^{-4}$.}\\
\end{tabular}
\end{table}

\end{document}